\documentclass[twocolumn,showpacs,pre,aps]{revtex4} 
\usepackage{graphicx} 
\usepackage{amsmath}

\DeclareMathOperator{\Erf}{Erf}

\begin{document}


  \title{Anomalous relaxation kinetics of biological lattice-ligand
    binding models}

  \author{Erwin Frey}
  \email{frey@hmi.de} 
        \homepage{http://www.hmi.de/bereiche/SF/SF5}
  \address{Abteilung Theorie, Hahn-Meitner-Institut,
    Glienickerstr. 100, 14109 Berlin, Germany}
  \address{Fachbereich Physik, Freie Universit{\"a}t Berlin,
    Arnimallee 14, 14195 Berlin, Germany} 
  
  \author{Andrej Vilfan}
        \email{av242@phy.cam.ac.uk}
        \homepage{http://www.tcm.phy.cam.ac.uk/~av242/}
  \address{TCM, Cavendish Laboratory, Madingley Road,
    Cambridge CB3 0HE, United Kingdom}

\date{5/2/2002}

% Abstract

\begin{abstract}
  We discuss theoretical models for the cooperative binding dynamics
  of ligands to substrates, such as dimeric motor proteins to
  microtubules or more extended macromolecules like tropomyosin to
  actin filaments.  We study the effects of steric constraints, size
  of ligands, binding rates and interaction between neighboring
  proteins on the binding dynamics and binding stoichiometry. Starting
  from an empty lattice the binding dynamics goes, quite generally,
  through several stages.  The first stage represents fast initial
  binding closely resembling the physics of random sequential
  adsorption processes.  Typically this initial process leaves the
  system in a metastable locked state with many small gaps between
  blocks of bound molecules.  In a second stage the gaps annihilate
  slowly as the ligands detach and reattach.  This results in an
  algebraic decay of the gap concentration and interesting scaling
  behavior. Upon identifying the gaps with particles we show that the
  dynamics in this regime can be explained by mapping it onto various
  reaction-diffusion models.  The final approach to equilibrium shows
  some interesting dynamic scaling properties. We also discuss the
  effect of cooperativity on the equilibrium stoichiometry, and their
  consequences for the interpretation of biochemical and image
  reconstruction results.
\end{abstract}


\maketitle

% Main text

\section{Introduction}

How does a system evolve towards its steady state? Sometimes the
answer is quite simple and the relaxation process is merely an
exponential decay. If the deviations from equilibrium are small
Onsager's regression hypothesis~\cite{onsager31} asserts that the
relaxation is governed by the same laws as the fluctuations in
equilibrium. This hypothesis certainly fails for systems with an
absorbing steady state such as simple models for diffusion-limited
chemical
reactions~\cite{torney83,toussaint83,racz85,peliti86,nielaba92}. Here
there are no fluctuations in the steady state but the approach towards
the absorbing state is critical in the sense that it exhibits slow
power-law decay and universal scaling behavior~\cite{cardyreview}.
Most of these models are chosen to be mathematically transparent
hoping that they still resemble some of the essential features of
actual systems occurring in nature. Unfortunately, experimentally
accessible systems where the above theoretical ideas can be tested
explicitly have remained rare to date.

\begin{figure*}[thbp]
\begin{center}
\begin{tabular}{l@{\hspace{2cm}}l}
\includegraphics{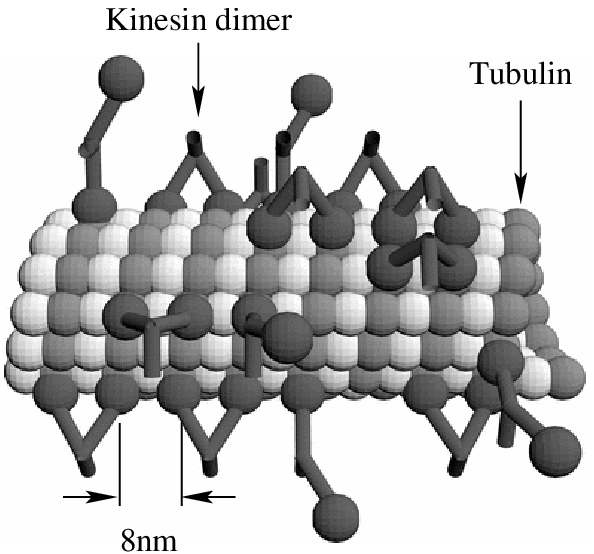}&
\includegraphics{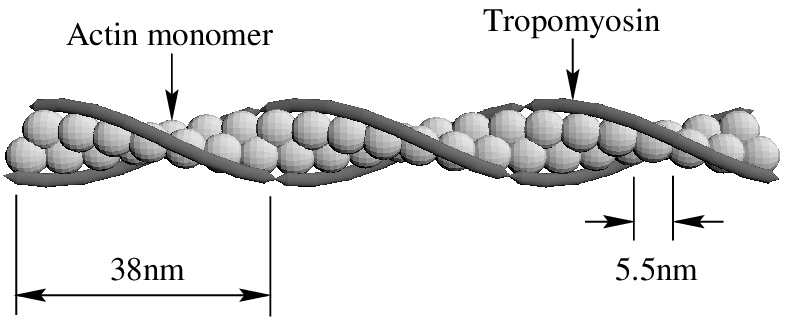}\\
a)&b)
\end{tabular}
\end{center}
\caption{\label{fig_reactions_a}
  a) Kinesin dimers can bind one head or both heads to a
  one-dimensional lattice (tubulin protofilament). Binding sites are
  located on $\beta$-tubulin subunits (dark) while $\alpha$-subunits
  (bright) are irrelevant for our study.\ b) Tropomyosin binds to an
  actin filament by occupying $7$ lattice sites.  There is a strong
  attractive interaction between the ends of bound tropomyosin
  molecules. \newline \tiny b) \copyright~2001, Biophysical Society \cite{vilfan2001c}}
\end{figure*}

In this contribution we discuss the kinetics of some macromolecular
assembly processes relevant for the formation of functional structures
in cells. In particular, we are interested in the dynamics of
ligand-substrate binding, where the substrate is a one- or
two-dimensional lattice and the ligands are dimers or oligomers.
Examples for such systems are illustrated in Fig.\ 
\ref{fig_reactions_a}.  Binding of dimeric myosin to actin
filaments~\cite{conibear98} can behave in a similar way as dimeric
kinesin on microtubules. Similar interactions of large supramolecular
biological polymers with protein ligands such as DNA with proteins or
viruses with antibodies are also a quite intensive area of research.
As will become clear in later chapters the kinetics of these systems
shows interesting anomalous dynamics which is closely related to the
mathematical models of simple diffusion-reaction systems discussed
above.

Fig.\ \ref{fig_reactions_a}a shows a schematic representation of a
{\it ``decoration experiment''}\/ where dimeric motor enzymes
(ligands) are deposited on their corresponding molecular track
(substrate)~\cite{mandelkow_hoenger99,hoenger00}.  For kinesin motors,
these tracks are microtubules, hollow cylinders usually consisting of
$13$ protofilaments, linear polymers composed of alternating $\alpha$-
and $\beta$-tubulin subunits. The kinesin binding sites are located on
the $\beta$ subunits (dark spheres) which form a helical (wound-up
rhombic) lattice with a longitudinal periodicity of $8\,{\rm nm}$.
Kinesin is a mechanochemical enzyme, which transforms (through an
isothermal stochastic process) the chemical energy obtained from the
hydrolysis of adenosine triphosphate (ATP) into motion along
microtubules.  Roughly speaking, kinesin has the following building
plan: two globular heads (also called motor domains) with the
functionality of "legs" are joined together by a coiled-coil region
into a tail which can bind to some cargo.  The typical size of these
proteins is in the range of several tens of nanometers.

Decoration techniques are usually performed in the absence of ATP
hydrolysis. Then the motor enzymes are ``passive'' ligands which bind
and unbind from their tubulin binding sites but do not actively move
along microtubules. These systems have traditionally been used in
biophysical chemistry to investigate the structure and the binding
properties of kinesin~\cite{mandelkow_hoenger99,hoenger00}, {\it
  i.e.}\ after waiting for the system to equilibrate the binding
stoichiometry is measured and the structure is determined by
cryo-electron microscopy followed by 3D image reconstruction. These
investigations are key elements in understanding the binding patterns
of kinesin motor domains under changing nucleotide conditions to get a
complete picture on the different conformational states, which are
involved in the kinesin walking mechanism~\cite{schief2001}.

In the following sections we present a theoretical analysis of dimer
adsorption-desorption kinetics with competing single and double bound
dimers. Such analysis is extremely important for a quantitative
analysis of decoration data~\cite{vilfan2001a}; a theoretical analysis
gives the binding stoichiometry in the equilibrium state in terms of
binding constants for the first and second head of the dimer molecule.
The dynamics of the approach to equilibrium is useful to estimate when
an experimental system can be considered as equilibrated.  Even more
importantly, time-resolved decoration experiments (e.g.\ by using
motor enzymes labelled by some fluorescent marker) combined with our
theoretical analysis could provide new information about reaction
rates which are to date not known completely.  Understanding the
kinetics of passive motors is undoubtedly a necessary prerequisite for
studying the more complicated case of active motors at high
densities~\cite{capello_preprint}.  The model is also interesting in
its own right since it contains some novel features of non-equilibrium
dynamics of dimer adsorption-desorption models.

Fig.\ \ref{fig_reactions_a}b shows a schematic representation of
tropomyosin binding to an actin filament~\cite{vilfan2001c}.  Actin
filaments are one of the major compontents of the cytoskeleton.  Like
microtubules, they contribute to the mechanical stability of the cell
and serve as tracks for molecular motors from the myosin family. This
function is especially pronounced in muscle cells where both actin and
myosin form filaments that can slide between each other and thereby
cause the muscle to contract.  Tropomyosin plays the key role in the
control of skeletal muscle contraction.  It binds to actin along its
binding sites for myosin motors.  When calcium ions are released as a
response to a nerve signal, they cause tropomyosin to shift laterally
thereby clearing the binding sites and allowing myosin to bind and
produce force.  Tropomyosin binding to action has several features
which make it an interesting model system of statistical physics; each
molecule covers seven actin monomers and interacts strongly with other
tropomyosin molecules. As a consequence gaps between bound molecules
can take a long time to heal. Although the gap dynamics has not yet
been measured experimentally, the amount of data gathered in other
kinetic studies provides plenty of information on the model parameters
and allows to make quantitative predictions about the relaxation time.
As it turns out \cite{vilfan2001c}, these relaxation times can be as
large as hours, which is short enough compared to the lifetime of
actin filaments in a muscle cell, but essential when planning
experiments with actin and tropomyosin assembled in vitro, especially
when studying the myosin regulation \cite{fraser95}.

The outline of this article is as follows.  In
Section~\ref{sec:decoration_model} we define the model for decoration
of microtubules with dimeric motors which can bind either with one or
with two heads as first introduced in Ref.~\cite{vilfan2001a}.  In
Section~\ref{sec:equil-stoich} we determine analytically the
equilibrium state of this model.  We study the dynamics of the model
in Section \ref{sec:dynamics}, which extends the results presented in
Ref.~\cite{vilfan2001b}.  Section~\ref{sec:dimer-model-with} studies
the dynamics of the two-dimensional model with interactions and
Section~\ref{sec:interacting-k-mers} the dynamics of the
one-dimensional $k$-mer model, which is a more generalized version of
the results described in Ref.~\cite{vilfan2001c}. In the final section
we give a summary and an outlook on future challenges in the field.

\section{Definition of the microtubule decoration model}
\label{sec:decoration_model}

\begin{figure}[thb]
  \centerline{\includegraphics{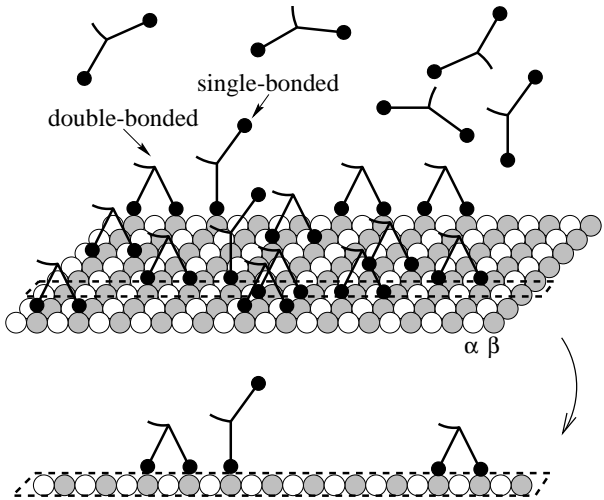}}
\caption{\label{fig_scheme3d_bw}
  Schematic representation of the binding of kinesin dimers to a
  tubulin sheet.  The binding sites are located on the $\beta$-tubulin
  subunits marked as grey balls. A dimer can attach either with one
  head to one binding site or with two heads on two adjacent sites
  along the same protofilament. Each binding site can be occupied by
  at most one kinesin head.  Neglecting interactions between kinesin
  molecules results in a one-dimensional model with dimers decorating
  individual protofilaments.}
\end{figure}

Our model describes the experimental situation as found in most
decoration assays. It starts with an empty tubulin sheet surrounded by
a solution of double-headed kinesin molecules. The kinesin dimers can
either attach with one or two heads onto binding sites located on
$\beta$-tubulin. There seems to be convincing evidence that kinesin
heads can bind on two adjacent binding sites only in longitudinal but
not in lateral direction~\cite{mandelkow_hoenger99,hoenger00}.  This
introduces a strong uniaxial anisotropy and distinguishes the
adsorption process of protein dimers from simple inorganic dimers. The
attached heads can also detach at some rate. A schematic
representation of the decoration process onto tubulin sheets is given
in Fig.\ \ref{fig_scheme3d_bw}.

\begin{figure}[thb]
  \centerline{\includegraphics{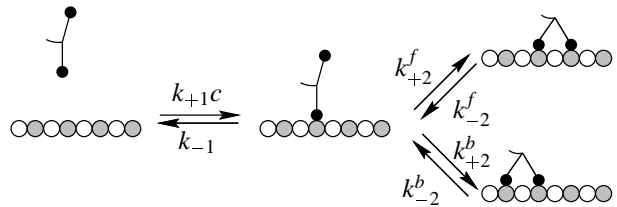}}
\caption{\label{fig_reactions_b} 
  Reaction scheme for all binding and unbinding processes. The rates
  for binding and unbinding of the first head are given by $k_{+1} \,
  c$ and $k_{-1}$, respectively. The second head may bind to an
  unoccupied site in front (f) of or behind (b) the already bound head
  with rates $k_{+2}^f$ and $k_{+2}^b$. The corresponding reverse
  process of detaching a front or rear head occurs at rates $k_{-2}^f$
  and $k_{-2}^b$. \newline \tiny \copyright~2001, Academic Press \cite{vilfan2001a}}
\end{figure}

\begin{figure}[thb]
  \centerline{\includegraphics{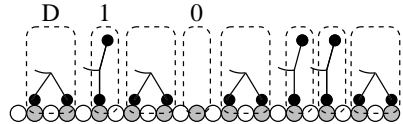}}
\caption{\label{fig_groups_bw} A configuration of bound kinesin
  molecules on a protofilament. The element $D$ represents a dimer with
  two bound heads, the element $1$ a dimer bound on one head and the
  element $0$ an empty lattice site (vacancy). \newline \tiny \copyright~2001, Academic Press \cite{vilfan2001a}}
\end{figure}

To begin with, let us take into account only steric (hard-core)
interactions and for now neglect nearest neighbour attractive
interactions. Then we are left with a one-dimensional problem of
kinesin dimers decorating a single protofilament (one-dimensional
lattice) as shown in Fig.\ \ref{fig_scheme3d_bw} and defined as
follows.\footnote{A Java applet with the simulation of our model can
  be found at {\tt http://www.ph.tum.de/\~{}avilfan/relax/}~.}
Kinesin is considered as a dimeric structure with its two heads
tethered together by some flexible joint. Hence each dimer (kinesin
protein) can bind one of its two heads (motor domains) to an empty
lattice site~\cite{vilfan2001a}.  The binding rate $k_{+1} \, c$ for
this process is proportional to the solution concentration $c$ of the
dimeric proteins. Successively, the dimer may either dissociate from
the protofilament with a rate $k_{-1}$ or also bind its second head to
an unoccupied site in front (f) of or behind (b) the already bound
head (with front we always refer to the direction pointing towards the
``+'' end of the microtubule, which is the direction of motion for
most motors from the kinesin family).  Since kinesin heads and
microtubules are both asymmetric structures the corresponding binding
rates $k_{+2}^f$ and $k_{+2}^b$ are in general different from each
other.  The reverse process of detaching a front or rear head occurs
at rates $k_{-2}^f$ and $k_{-2}^b$. A reaction scheme with all
possible processes and their corresponding rate constants is shown in
Fig.\ \ref{fig_reactions_b}.
In typical decoration experiments there is no external energy source,
{\it i.e.}\/ no ATP hydrolysis. Then the binding rates are not all
independent of each other, but detailed balance dictates that the
ratio of on- and off-rates has to equal the equilibrium binding
constants
\begin{equation}
K_1 = \frac{k_{+1}}{k_{-1}} \, , \quad \text{and} \quad
K_2 = \frac{k_{+2}^b}{k_{-2}^b} = \frac{k_{+2}^f}{k_{-2}^f} \,.
\end{equation}
A particular coverage of the lattice is described as a sequence of
dimers bound with both heads ($D$), one head only ($1$) and empty
sites ($0$) (see Fig.~\ref{fig_groups_bw}). 
We denote the probabilities to find a certain lattice
site in one of these states by $2 n_D$, $n_1$ and $n_0$, respectively.
Of course, normalization of the probabilities requires $ n_0 + n_1 + 2
n_D=1$.

\subsection{Particle-hole symmetry}
\label{sec_symmetry}

\begin{figure}[thb]
\begin{center}
  \includegraphics{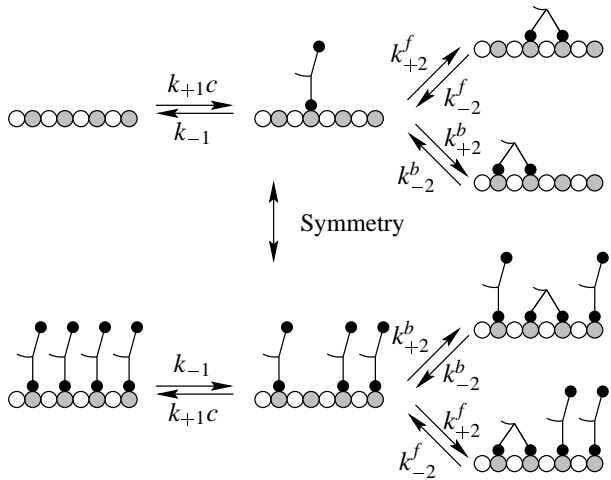}
\end{center}
\caption{\label{fig_symmetry} The one-dimensional dimer decoration 
  model is invariant upon exchanging empty lattice sites $(0)$ with
  single-bonded dimers $(1)$ and the reaction rates according to
  $(k_{+1}c,k_{2\pm}^f) \leftrightarrow (k_{-1},k_{2\pm}^b)$.}
\end{figure}

Important information about the steady state and the dynamics of the
system can be gained already by exploiting the symmetries of the
kinetic process.  Fig.\ \ref{fig_symmetry} reveals a ``particle-hole''
symmetry by showing a transformation that maps the system onto one
with the same reaction scheme, albeit with transformed kinetic
constants. The details of the transformation are listed in Table
\ref{tab_symmetries}.

The ``particle-hole'' exchange operation replaces dimers bound on a
single head (``particles'') by vacant sites (``holes'') and vice
versa. Dimers bound with both heads are kept invariant under the
transformation. To obtain a system with an equivalent reaction scheme
the transition rates have to be transformed as well. This is achieved
by exchanging the attachment and detachment rate of the first head,
$k_{+1} c \leftrightarrow k_{-1}$, and simultaneously switching the
forward and backward binding rates of the second head, $k_{2\pm}^f
\leftrightarrow k_{2\pm}^b$. As a consequence of this symmetry
operations the equilibrium constant $K_1 c$ is replaced by its
reciprocal value while the binding constant of the second head $K_2$
is left unchanged.
\begin{table*}[thb]
\caption{The dynamics of the one-dimensional dimer decoration model 
 is invariant under the following ``particle-hole'' transformation.}
\label{tab_symmetries}
\begin{tabular}{l|ccc|cccccc|cc}
&\multicolumn{3}{c|}{state}&
\multicolumn{6}{c|}{kinetic constants}&
\multicolumn{2}{p{2.5cm}}{equilibrium constants}\\
\hline
Original & 0 & 1 & D & $k_{+1}c$ &   $k_{-1}$ &  $k_{+2}^f$ & $k_{+2}^b$ &
$k_{-2}^f$ & $k_{-2}^b$ & $K_1 c$ & $K_2$ \\
Transformed & 1 & 0 & D &  $k_{-1}$ & $k_{+1}c$ &  $k_{+2}^b$ & $k_{+2}^f$ &
$k_{-2}^b$ & $k_{-2}^f$ & $1/(K_1 c)$ & $K_2$
\end{tabular}
\end{table*}

This symmetry has immediate consequences on the equilibrium
stoichiometry.  Since the coverage in the steady state is solely a
function of the quantities $K_1 c$ and $K_2$ this symmetry implies
that the mean number of dimers attached with both heads (which are
mapped onto themselves) $n_D (K_1 c, K_2)$ is invariant upon
interchanging the attachment and detachment rates of the first head,
\begin{equation}
n_D(1/K_1c,K_2) = n_D (K_1c,K_2)
\end{equation}
Similarly the mean total number of bound heads per lattice site
(binding stoichiometry), $\nu = 2(n_1 + n_D)$, obeys the symmetry
relation
\begin{equation}
\nu(1/K_1c,K_2)=2-\nu(K_1c,K_2)\;.
\end{equation}
The symmetries can be best seen in a logarithmic-linear plot as
shown in Fig.\ \ref{fig_curves}.  From these relations we already
conclude that the stoichiometry of dimers bound with both heads $n_D$
reaches its maximum at $K_1c=1$ and that the total stoichiometry is
$\nu=1$ at that point.

\subsection{Experimental parameter values}

The kinetic constants for the binding of kinesin on microtubules have
been determined by several groups
\cite{hackney94,ma95,moyer98,mandelkow98}.  The binding constants for
the first head in the presence of ATP and low ionic strength have the
values \cite{moyer98} $k_{+1}=20 \, \mu{\rm M}^{-1}{\rm s}^{-1}$ and
$k_{-1}=25 \, {\rm s}^{-1}$, leading to $K_1=0.8 \,\mu {\rm M}^{-1}$.
$k_{-1}$ is much smaller in the presence of AMP-PNP or in the absence
of a nucleotide, about $0.01 \, {\rm s}^{-1}$ \cite{hancock99}.  The
binding constant $K_2$ can be estimated indirectly as the ratio
between the detachment rate of the monomeric and the dimeric kinesin
and has values between 2.7 (in the presence of ATP) and 20 (without a
nucleotide) \cite{hancock99}.  These values show that the model
parameters depend strongly on the chemical conditions and the type of
motor protein used in the experiment. A theoretical investigation
allows for a systematic analysis and detailed classification of all
the different regimes of binding kinetics in such a broad range of
parameter values.

\section{Equilibrium stoichiometry}
\label{sec:equil-stoich}

In this section we review results obtained for the equilibrium
stoichiometry of the dimer binding model \cite{vilfan2001a}. In this
section we will discuss the one-dimensional model where some
short-range interaction between the dimers is taken into account;
two-dimensional models are studied in Sec.\
\ref{sec:dimer-model-with}.

\subsection{Analytical solution for the binding stoichiometry}

\begin{figure}[thbp]
\begin{center}
\includegraphics{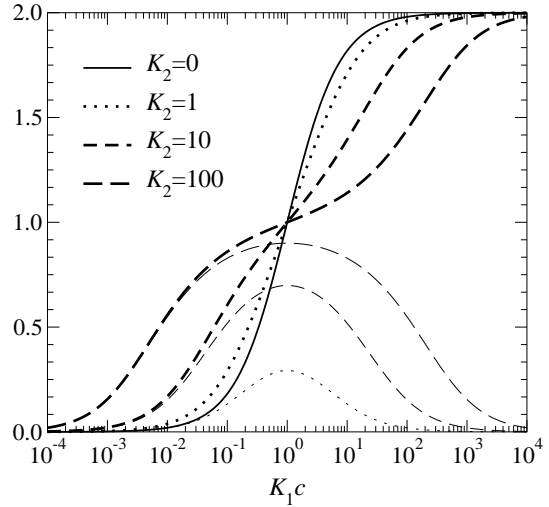}
\end{center}
\caption{\label{fig_curves}Binding stoichiometry $\nu$  as a function of the
  solution concentration $c$ and the constant $K_2$. The thick curves
  show the total number of heads per lattice site $\nu$ (\ref{eq8}).
  The thin curves show the fraction of binding sites occupied with
  double-attached dimers $2 n_D$ (\ref{eq8a}).}
\end{figure}

For the one-dimensional model the value of the mean occupation numbers
in the steady state can quite easily be determined upon using detailed
balance and the fact that the dimers have only a hard-core
interaction. Because we have non-interacting dimers the probability to
find a certain sequence of $0$'s, $1$'s and $D$'s ({\it e.g.}\ 
``0,1,D,D,1,D'', see Fig.\ \ref{fig_groups_bw}) has to be invariant
against permutations of these states. In such a random sequence the
probability $p_i$ to find a particular state $0$, $1$ or $D$ at a
certain site $i$ is given by
\begin{equation}
p_i = \frac{n_i}{n_0 + n_1 + n_D} \,.
\end{equation}  
Detailed balance requires that for each pair of possible
configurations, their respective probabilities are in the same ratio
as the transition rates between them. Hence the ratio of probabilities
to find a sequence with a $1$ or $0$ at a certain place is
\begin{equation}
\label{eq_balance1}
\frac{p_1}{p_0} = \frac{k_{+1}c}{k_{-1}}=K_1c \,.
\end{equation} 
Similarly, we get for transitions between $D$ and $01$: 
\begin{equation}
\frac{p_D}{p_0 p_1} = \frac{k_{+2}^{f}}{k_{-2}^{f}} 
                    = \frac{k_{+2}^{b}}{k_{-2}^{b}} = K_2 \,.  
\end{equation}
These two equations, together with the normalization condition,
uniquely determine the values $n_0$, $n_1$ and $n_D$.  The number of
dimers bound with both heads is given by
\begin{equation}
\label{eq8a}
n_D = \frac12 -  \frac12 \left(\frac{4 K_1
K_2 c}{(1+K_1 c)^2}+1\right)^{-\frac12}
\end{equation}
and the number of dimers bound with one head
\begin{equation}
\label{eq8b}
n_1 = \frac{K_1c}{K_1c+1} \left(\frac{4 K_1
K_2 c}{(1+K_1 c)^2}+1\right)^{-\frac12}
\end{equation}
The stoichiometry, {\it i.e.}\/ the total number of heads per binding
site $\nu = 2(n_1+n_D)$ then reads
\begin{equation}
\label{eq8}
  \nu = 1+ \frac{K_1 c-1}{K_1 c+1}
  \left(\frac{4 K_1 K_2 c}{(1+K_1 c)^2}+1\right)^{-\frac12}\;.
\end{equation}
The number of dimers bound with both heads per lattice site reaches
its maximum $n_D^{\rm max}=(1 - 1 /\sqrt{K_2+1})/2$ for $K_1 c=1$. For
an illustration of all these equations see Fig.\ \ref{fig_curves}.

\subsection{Alternative derivation of the binding stoichiometry}

Here we present an alternative derivation of the binding stoichiometry
which has the additional benefit that it can also be used for
two-dimensional lattices where dimers can bind in either direction.
The idea is to relate the equilibrium stoichiometry of the flexible
dimer model to that of a stiff dimer model -- a model where dimers can
only bind as whole.  Once we have specified the mapping one can
determine the stoichiometry of the general model from the
stoichiometry of the stiff dimer model in one dimension which has been
known for quite a long time \cite{hill57} already and represents a special case
of the solution by McGhee and von Hippel \cite{mcghee74}
\begin{equation}
\label{eq_stoch_stiff}
n=\frac{1}{2} \left( 1 - \frac{1}{\sqrt{4 K +1}} \right)\;.
\end{equation}
Note, that in this notation the binding constant $K$ contains the
concentration $c$.  In equilibrium, a mapping between the stiff and
flexible dimer model is possible because of detailed balance and
because transitions between states $0$ and $1$ are uncorrelated with
the surrounding configuration.  Let's denote the equilibrium binding
constant, the dimer and vacancy density in the stiff dimer model by
$K$, $\bar n_D$ and $\bar n_0$, respectively.  Then, detailed balance
gives the following relation between these densities and the
equilibrium binding constant, $K=\bar n_D / \bar n_0^2$. In the same
way detailed balance implies $K_2=n_D / (n_0 n_1)$ and $K_1c=n_1/n_0$
for the general dimer model. The absence of correlations allows us to
group together the states $1$ and $0$ of the general model into a
single state which we identify with the vacant state $\bar 0$ of the
stiff dimer model.  This gives the mapping as summarized in Table
\ref{tab_mapping}.

\begin{table}[thbp]
\caption{Mapping between the equilibrium states of the stiff and the 
  flexible dimer model.}
  \label{tab_mapping}
\begin{center}
\begin{tabular}{l|ll}
&Stiff & Flexible\\
\hline
Densities & $\bar n_D$, \, $\bar n_0$ & $n_D$,  \, $n_0+n_1$\\
Binding constants& $K$ & $K_1c$, $K_2$\\
Equilibrium condition& $K=\frac{\bar n_D}{\bar n_0^2}$ 
& $K_1c=\frac{n_1}{n_0}$, $K_2=\frac{n_D}{n_0 n_1}$\\
\end{tabular}
\end{center}
\end{table}

From this table we can infer that the number of fully bound dimers per
lattice site in the flexible dimer model is given as
\begin{equation}
  \label{eq_map1}
  n_D(K_1 c, K_2) = \bar n_D(K)
                  = \bar n_D\left( \frac{K_2 K_1 c}{(1+K_1c)^2}\right)
                  \; ,
\end{equation}
and the number of dimers bound with one head as
\begin{multline}
  \label{eq_map2}
  n_1(K_1 c, K_2)=\frac{K_1 c}{1+K_1 c} \bar n_0 (K) =\\= 
  \frac{K_1 c}{1+K_1 c}\left( 1-\frac12  \bar n_D\left( \frac{K_2 K_1 c}{(1+K_1c)^2}
  \right) \right)\;. 
\end{multline}
In the one-dimensional case we can insert Eq.\ (\ref{eq_stoch_stiff})
and reproduce Eqs.\ (\ref{eq8a}) and (\ref{eq8b}). Note that the
mapping reduces the number of parameters by one leading to a quite
significant simplification. Though the stiff dimer model has not yet
been solved analytically in two dimensions, the model with a single
parameter can easily be solved by Monte-Carlo simulation (e.g.,
\cite{phares93b}). The entropy of a two-dimensional lattice fully
covered with dimers is even known
exactly~\cite{kasteleyn61,temperley61}.

\section{Dynamics}
\label{sec:dynamics}

In this section we will learn that the relaxation towards the
equilibrium state described in the previous section is by no means a
simple process characterized by a single time scale but actually shows
quite interesting anomalous kinetics. We find different scenarios
depending on the value of the equilibrium binding constants $K_2$ of
the second head.  If $K_2 \lesssim 1$ there are comparatively few
double-bonded dimers in the steady state.  Consequently cooperative
effects and correlations are less important implying that the
equilibration process is exponential with the time scale given by the
rates $k_{+1}c$ and $k_{-1}$. The dynamics changes drastically if the
second head is very likely to bind, $K_2 \gg 1$. Then, the equilibrium
state will consist mainly of dimers bound with both heads aligned with
twice the period of the lattice.  Their positions are correlated at a
length scale given by the average distance between defects in the
periodic pattern (either vacancies or single-bonded dimers).  As we
will detail in this section these correlations lead to a dynamics
which is slowed down drastically as compared to the kinetics of
individual dimers.

In the following we will mainly concentrate on the case $K_1 c \ll 1$,
to which we refer as the ``stiff dimer model'', reflecting the fact
that dimers are only likely to be found in the state with both heads
bound.  In this limit we can discuss the essential features of our
model, which include the anomalous relaxation kinetics.  Because of
the particle-hole symmetry (Sect.~\ref{sec_symmetry}) which transforms
$K_1 c \to 1/(K_1 c)$, the results apply equally well for $K_1 c \gg
1$.  The other interesting limiting case is the situation with $K_2
\gg 1$ and $K_1 c \approx 1$. In this limit the dynamics is
essentially similar to the case $K_1 c \ll 1$, but more complicated in
detail because we have to consider all three states, {\it i.e.}\/
empty sites, dimers bound with one head and dimers bound with both
heads.

\subsection{The stiff dimer limit}
\label{sec:stiff-dimer-limit}

It turns out that most interesting features of the model can already
be discussed if we restrict ourselves to the limit
\begin{equation}
\label{eq_stiff_condition}
K_2\gg 1, \qquad K_1 c \ll 1
\end{equation}
with $K=K_1 K_2 c$ fixed at a value which allows the equilibrium state
to consist mainly of dimers bound with both heads.  The first
condition of Eq.\ (\ref{eq_stiff_condition}) states that the second
head is unlikely to be found in the unbound state if it has a place
where it can bind.  The second condition implies that isolated
single-site vacancies are unlikely to be occupied with single-bonded
dimers.  Therefore, our model reduces to a dimer
deposition-evaporation model where the dimers can only bind and unbind
with both heads at the same time.  We will refer to this model as
``stiff dimer model'' in the following.  The vacancy concentration
$n_0$ in the steady state, given as $1-n_1-2n_D$ in Eq.\ (\ref{eq8a})
and (\ref{eq8b}) then simplifies to
\begin{equation}
\label{eq_gap_dens_stiff}
n_0=1/\sqrt{1+4K}\;.
\end{equation}

The stiff dimer model describes dimers that bind at once to a pair of
empty lattice sites and is characterized by the effective attachment
rate of a whole dimer to an empty pair of sites, $k_+c$, an effective
detachment rate $k_-$, and a rate with which a dimer can move
diffusively without detaching, $r_d$.  Depending on whether the
diffusion or the detachment rate is larger, we will distinguish between
the \emph{diffusive} and the \emph{non-diffusive} case.  The stiff
dimer model is similar to conventional dimer deposition models
\cite{privman92}, however with the difference that the attachment and
detachment constants $k_+$ and $k_-$ also depend on whether the
neighboring sites are occupied or not (but we will show that this
dependence becomes irrelevant in the non-diffusive case).

\subsubsection{Mapping between the general model and the stiff dimer model}

In the following we describe how the kinetics of the stiff dimer model
follows from our general dimer model.  In the limit we are
considering, the detachment of a single-bound dimer is always much
faster than the attachment of a new one ($k_{-1}\gg k_{+1}c$,
following from $K_1 c \ll 1$) and the attachment of the second head
much faster than its detachment ($k_{+2}^{f,b}\gg k_{-2}^{f,b}$,
following from $K_2 \gg 1$).  We do not put any other limitations on
the reaction rates for now.

\begin{figure}[thbp]
\centering{\includegraphics{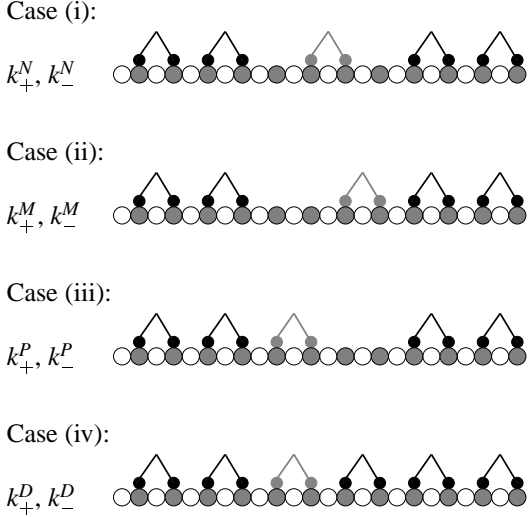}}
\caption{Four cases for the binding of a new dimer (grey) 
  in a gap between already bound ones (black).  The effective binding
  rates are $k_+^{N,M,P,D}$ and the unbinding rates $k_-^{N,M,P,D}$.}
  \label{fig_att_cases}
\end{figure}

In a general treatment, the attachment and detachment rates also
depend on the occupancy of the neighboring sites.  We therefore have
to distinguish between four cases: both neighbors empty, front
neighbor occupied, rear neighbor occupied and both neighbors
occupied.
\begin{itemize}
\item \textit{Case (i): Both neighbors empty}.  A dimer attaches with
  its first head at the rate $k_{+1}c$.  Afterwards, the second head
  can attach at the total rate $k_{+2}^f+k_{+2}^b$, while the first
  head can detach with rate $k_{-1}$.  This leads to the total
  attachment rate per lattice site
\begin{equation}
\label{eq_a_case1}
k^N_+c =\frac{k_{+1}c(k_{+2}^f+k_{+2}^b)}{k_{-1}+k_{+2}^f+k_{+2}^b}\;.
\end{equation}
The process of detachment starts with the detachment of the front or
rear head. The detachment gets completed if the other head detaches
too (rate $k_{-1}$) and fails if the detached head reattaches again
(rate $k_{+2}^f+k_{+2}^b$).  The detachment rate then becomes
\begin{equation}
\label{eq_d_case1}
k^N_- =\frac{k_{-1}(k_{-2}^f+k_{-2}^b)}{k_{-1}+k_{+2}^f+k_{+2}^b}\;.
\end{equation}

\item \textit{Case (ii): Front neighbor occupied}.  There are two
  pathways that lead to binding next to an occupied site.  The first
  head can either attach on the site next to the occupied one or one
  site further away, both at the rate $k_{+1}c$. In the first case the
  probability that the second head will attach before the first one
  detaches is $k_{+2}^b/(k_{-1}+k_{+2}^b)$.  In the second case the
  probability that the second head attaches in front of the first one
  is $k_{+2}^f/(k_{-1}+k_{+2}^f+k_{+2}^b)$.  These two terms taken
  together give the attachment rate to the given pair of sites
\begin{equation}
\label{eq_a_case2}
k^M_+c =k_{+1}c \left(
\frac{k_{+2}^b}{k_{-1}+k_{+2}^b} + \frac{k_{+2}^f}{k_{-1}+k_{+2}^f+k_{+2}^b}
\right)\;.
\end{equation}
If the process of detachment starts with detaching the front head
(rate $k_{-2}^f$), the probability that the whole dimer will detach
afterwards is $k_{-1}/(k_{-1}+k_{+2}^f+k_{+2}^b)$.  If it starts with
detaching the rear head (rate $k_{-2}^b$), this probability is
$k_{-1}/(k_{-1}+k_{+2}^b)$.  This gives the total detachment rate
\begin{equation}
\label{eq_d_case2}
k^M_- =k_{-1} \left(
\frac{k_{-2}^b}{k_{-1}+k_{+2}^b} + \frac{k_{-2}^f}{k_{-1}+k_{+2}^f+k_{+2}^b}
\right)\;.
\end{equation}

\item \textit{Case (iii): Rear neighbor occupied}.  This case is
  analogous to the previous one, except that the indices $f$ and $b$
  have to be interchanged.  The attachment rate becomes
\begin{equation}
\label{eq_a_case3}
k^P_+c =k_{+1}c \left(
\frac{k_{+2}^b}{k_{-1}+k_{+2}^f+k_{+2}^b} + \frac{k_{+2}^f}{k_{-1}+k_{+2}^f}
\right)
\end{equation}
and the detachment rate
\begin{equation}
\label{eq_d_case3}
k^P_- =k_{-1} \left(
\frac{k_{-2}^b}{k_{-1}+k_{+2}^f+k_{+2}^b} + \frac{k_{-2}^f}{k_{-1}+k_{+2}^f}
\right)\;.
\end{equation}

\item \textit{Case (iv): Both neighbors occupied}.  In this case the dimer
  binds into a gap of exactly two sites. If the first head attaches to the
  front site in the gap (rate $k_{+1}c$), the probability for the second one to
  attach before the first one detaches is $k_{+2}^b/(k_{-1}+k_{+2}^b)$.
  Together with the analogous case in which the first head attaches to the rear
  site inside the vacancy, the total attachment rate becomes
\begin{equation}
\label{eq_a_case4}
k^D_+c =k_{+1}c \left(
\frac{k_{+2}^b}{k_{-1}+k_{+2}^b} + \frac{k_{+2}^f}{k_{-1}+k_{+2}^f}
\right) \; .
\end{equation}
After the front head has detached (rate $k_{-2}^f$), the probability
that the whole dimer will follow is $k_{-1}/(k_{-1}+k_{+2}^f)$.
Adding the pathway starting with the detachment of the rear head gives
the detachment rate
\begin{equation}
\label{eq_d_case4}
k^D_- =k_{-1} \left(
\frac{k_{-2}^b}{k_{-1}+k_{+2}^b} + \frac{k_{-2}^f}{k_{-1}+k_{+2}^f}
\right)\;.
\end{equation}
\end{itemize}
Of course, all these rates obey detailed balance, which states that
\begin{equation}
\frac{k_+c}{k_-}=K=K_1 K_2 c\;.
\end{equation}

Processes in which one head detaches on one side and subsequently
attaches on the other side also lead to an {\em explicit} diffusion of
attached dimers. A diffusive step forwards occurs if the rear head
detaches (rate $k_{-2}^b$) and reattaches on the front side
(probability $k_{+2}^f/(k_{-1}+k_{+2}^f+k_{+2}^b)$).  The hopping rate
(which is, of course, equal for backward steps) then reads
\begin{equation}
\label{d_dif}
r_d=\frac{k_{+2}^f k_{+2}^b}{K_2(k_{-1}+k_{+2}^f+k_{+2}^b)}\;.
\end{equation}

So far we have shown that in the limit $K_2 \gg 1$ our model can be
mapped onto a simple dimer model.  But, depending on parameters, this
can happen in two qualitatively different ways.  First, if the
detachment rate is smaller than the diffusion rate $k_- \ll r_d$,
which is equivalent to the condition
\begin{equation}
\label{eq_diffusive_condition}
k_{-1} \ll  \frac{1}{\frac{1}{k_{+2}^f}+\frac{1}{k_{+2}^b}}\;,
\end{equation}
we get the \textit{diffusive limit}. In the opposite limit, a dimer is
unlikely to make a diffusive step before it detaches and we call that
the \textit{non-diffusive limit}.  In the non-diffusive limit, the
effective dimer attachment ($k_+c$) and detachment ($k_-$) rates
become equal in all four cases, {\it i.e.}\/ they become independent
of the surroundings of a dimer.

To summarize, the stiff dimer model in the non-diffusive limit, to
which we will concentrate in the following, describes dimers that can
bind to any free pair of lattice sites with equal rate, namely $k_+c$
and can detach with a rate $k_-$.  The dimer cannot move along the
lattice and it always stays joined (cannot regroup its two parts with
parts of other dimers).  Although dimers cannot diffuse directly, in
the case of high lattice coverage we can still introduce an effective
diffusion which describes dimers near a gap detaching and reattaching
on the other side of the gap.  This simplified model still captures
all the essential aspects of the non-equilibrium dynamics.

\subsubsection{Related dimer models}

Similar dimer adsorption models have been studied
previously~\cite{privman92,barma93,stinchcombe93,eisenberg97}.
Privman and Nielaba~\cite{privman92} studied the effect of diffusion
on the dimer deposition process (neglecting dimer desorption). There
are several key differences to our model, the most important of which
is that diffusion without detachment results in a 100\,\% saturation
coverage, whereas a model with detachment leads to a limiting coverage
whose value depends on the binding constants of the first and second
head (see Eq.~\ref{eq8}).  This also has important implication on the
dynamics as discussed below. For example, as a consequence of a finite
coverage the final approach to equilibrium is not a power law but
exponential and there are, in addition, interesting temporal
correlations in the fluctuations in the steady state.

Stinchcombe and coworkers~\cite{barma93,stinchcombe93} studied the
effect of detachment on the adsorption kinetics but allowed for
regrouping of attached dimer molecules (two monomers that belonged to
different dimers during attachment can form a dimer and detach
together).  While such processes are allowed for some types of
inorganic dimers, they are certainly forbidden for dimer proteins like
kinesin where the linkage between its two heads is virtually
unbreakable. If regrouping is allowed the {\em steady state}
auto-correlation functions for the dimer density shows an interesting
power-law decay $\propto t^{-1/2}$~\cite{barma93,stinchcombe93}.  This
behavior can be directly linked to the gapless spectrum in an
associated spin model~\cite{stinchcombe93}.  If regrouping is
forbidden this power-law decay is lost (and becomes an exponential to
leading order) due to the permanent linkage between the two heads of
the dimer.  Intuitively this may be understood as follows: only if
regrouping of dimers is allowed are there locally jammed
configurations (N\'eel-like states, ``101010'', with alternating
occupied and unoccupied sites in which neither attachment nor
detachment of dimers is possible) in the final steady state which slow
down the dynamics.  The difference between the occupation numbers on
even and odd sites then represents a conserved quantity slowing down
the dynamics.  In our model, the conservation law is trivial since the
difference between occupation numbers on even and odd sites always
disappears.

In addition, we will see in Sec.\ \ref{subsec_steadystate} that the
autocorrelation functions of the dimer and the vacancy occupation
number show strong differences in shape and typical times scales of
relaxation.

\subsubsection{Numerical analysis of the non-diffusive stiff dimer model}

\begin{figure}[thb]
  \centerline{\includegraphics*{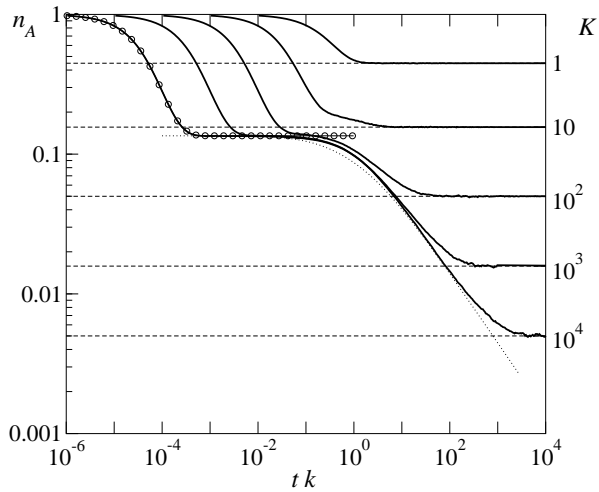}}
\caption{\label{fig_relax} 
  Vacancy concentration in the non-diffusive stiff dimer model as a
  function of time $t$ for $K=k_+c/k_-=1,\, ... \, , 10000$.  Dashed
  lines show the steady state concentration (\ref{eq_gap_dens_stiff})
  for a given $K$. The line with circles shows the short-time limit
  for $K=10000$ (\ref{eq_rsa1}), and the dashed line the result of the
  reaction-diffusion model (Eq.\ \ref{eq_rd}). }
\end{figure}

To study the kinetics we choose the initial condition as typically
used in an experiment, namely an empty lattice.  Figure
\ref{fig_relax} shows simulation data for the average vacancy
concentration as a function of time for a set of binding constants
$K=k_+c/k_-\equiv K_1 K_2 c$.  We find qualitatively very different
approaches to the final steady state depending on the value of the
binding constant $K$.  For $K \ll 1$, where the off-rate $k_-$ is much
larger than the on-rate $k_+ c$, there is no crowding on the lattice
and the dimeric nature of the molecules does not affect the approach
to equilibrium. Hence one gets, like for monomers, an exponential
behavior with a decay rate $k_-$.  In the opposite limit, $K \gg 1$,
there is a pronounced {\em two-stage relaxation} towards the steady
state.  The vacancy concentration as a function of time reveals {\em
  four regimes}, an initial attachment phase, followed by an
intermediate plateau, then a power-law decay and finally an
exponential approach towards equilibrium.

\paragraph{Initial binding}
At short time scales, $t \ll k_-^{-1}$ and $t \ll r_d^{-1}$, when only
deposition processes are frequent but detachment or diffusion
processes are still very unlikely, the kinetics of the model belongs
to the class of problems frequently referred to as \textit{random
  sequential adsorption} (RSA) models (reviewed by
Evans~\cite{evans93} and Talbot \emph{et al.}~\cite{talbot00}).  RSA
Models have among others been used to describe the chemisorption of
inorganic dimers like $\mathrm{O}_2$ on crystal surfaces, the binding
of reagents to organic polymers or chemical reactions on the polymer
chain.

The deposition of stiff dimers has first been studied by
Flory~\cite{flory39}.  Later, it was pointed out by Page~\cite{page59}
(see also a note by Downton~\cite{downton61}) that the final
distribution depends on whether dimers bind with both ends in parallel
(``\emph{standard model}'') or with one end first (``\emph{end-on
  model}'').  A more recent article addressing this point and giving
analytical solutions for both cases was published by Nord and
Evans~\cite{nord90}.  In the ``standard model'' the dimers bind with
equal rates to all free pairs of lattice sites. In the ``end-on'' the
first end of a dimer binds with equal rates to any free site.  The
second end then binds to either side if there is space, or to the only
available side if the other one is occupied.  This has the consequence
that the probability that a dimer binds to either end of a gap is
$1.5$ times higher than that it binds to a certain pair of sites in
the middle of the gap.  This obviously leads to some clustering of
dimers reduces the number of gaps that remain in the end.

It is even possible to exactly calculate the time-dependence of the
lattice coverage~\cite{mcquistan68}.  In the ``standard model'' the
vacancy density during the RSA phase follows
\cite{mcquistan68,evans93}
\begin{equation}
\label{eq_rsa1}
n_0(t)=\exp\left(-2+2e^{-k_+c t}\right)\;.
\end{equation}
The vacancy concentration locks at an intermediate plateau 
\begin{equation}
\label{eq_rsa2}
n_0^{\rm plateau} = e^{-2}\approx 0.1353
\end{equation}
This Flory plateau represents a configuration in which all remaining
vacancies are isolated, causing the system to be unable to accommodate
for the deposition of additional dimers.  In the ``end-on'' model the
plateau vacancy concentration is
\begin{equation}
\label{eq_rsa3}
n_0^{\rm plateau}=\sqrt{2\pi e}(\Erf(\sqrt2)-\Erf(1/\sqrt 2)) -1 \approx 0.1233\;,
\end{equation}
slightly smaller than in the non-diffusive limit. 

When applying the RSA results to our model, we have to distinguish
between the diffusive and the non-diffusive case.  In the
\emph{non-diffusive} case, the binding rate at all pairs of sites,
regardless whether their neighboring sites are empty
(\ref{eq_a_case1}), one of them (\ref{eq_a_case2}, \ref{eq_a_case3})
or both (\ref{eq_a_case4}) are occupied, is equal.  This corresponds
to the ``standard model''.  The time-dependent gap density is given by
(\ref{eq_rsa1}) and its plateau value by (\ref{eq_rsa2}).

In the \emph{diffusive} case, the binding rates on both ends of an
interval, (\ref{eq_a_case2}) and (\ref{eq_a_case3}), add up to a rate
three times as large as the binding rate in the middle of an interval
(\ref{eq_a_case1}), while the attachment rate for a pair of sites with
both neighbors occupied (\ref{eq_a_case4}) is twice as large.
Therefore, the RSA phase in the diffusive limit corresponds to that of
the ``end-on model''.  The plateau gap density is then given by
(\ref{eq_rsa3}).  In both cases, the final configuration is
independent of the asymmetry in the binding rates.

\begin{figure}
\centerline{\includegraphics{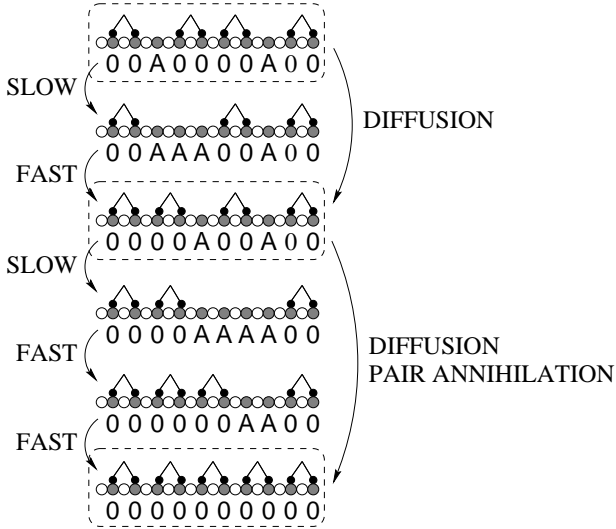}}
\caption{\label{fig_cg}Time evolution of a state, consisting of 
  attachment (fast) and detachment (slow) events.  The coarse-grained
  interpretation includes only long-living states (in boxes).  The
  effective steps include diffusion, pair annihilation and pair
  creation (not shown). \newline \tiny \copyright~2001, EDP Sciences \cite{vilfan2001b}}
\end{figure}

\paragraph{Power-law regime}
\label{sec:power-law-regime}
The secondary relaxation process towards the final equilibrium state
is much slower than the initial random sequential adsorption process.
Starting out of a jammed configuration in the Flory plateau the
dynamics shows a broad time domain with a power-law $\propto t^{-1/2}$
instead of a simple exponential decay. Similar multi-stage relaxation
processes have been observed in dimer models with diffusion but no
detachment \cite{privman92}, the key difference being that the
detachment process implies that the steady state has a finite vacancy
density and the final approach to the steady state remains not a power
law but becomes exponential. There are also interesting similarities.
In particular, in both models a large portion of the final approach to
the steady state is mediated by the annihilation of vacancies. This
behavior can be explained by introducing a particle representation in
the following way (analog to the adsorption-diffusion model
\cite{privman92}).  We denote each vacancy on the lattice as a
``particle'' $A$, and each bound dimer as an inert state ($00$).  The
detachment of a dimer then corresponds to a pair creation process $00
\to AA$, and the decoration process to pair annihilation $AA \to 00$.
Since we consider the limit $K \gg 1$, states with two vacancies $A$
on neighboring sites have a very short lifetime.  We may thus
introduce a coarse-grained model by eliminating these states (see
Fig.\ \ref{fig_cg}).

\begin{table*}[thbp]
\caption{The processes in the $A+A\leftrightarrow 0$ model onto which 
the stiff dimer model can be mapped in the limit $K\gg 1$.}
\label{tab_rates}
  \begin{tabular}{lllll}
&    Process & Rate & Branching & Total\\
&            & (1)  & prob. (2) & rate\\
    \hline
Hopping  &  $A00\stackrel{(1)}\to AAA\stackrel{(2)}\to A00$ 
         & $k_-$ 
         & $\frac12$ 
         & $\frac{k_-}{2}$\\
Annihilation & $A00A \stackrel{(1)}\to AAAA \stackrel{(2)}\to 
               \left\{\begin{array}{c}00AA\\AA00 \end{array}\right\}\to 0000$ 
             & $k_-$ 
             & $\frac23$ 
             & $\frac{2k_-}{3}$\\
Creation & $0000\stackrel{(1)}\to \left\{\begin{array}{c}00AA\\AA00
          \end{array}\right\} \stackrel{(2)}\to AAAA \stackrel{(2)}\to A00A$ 
         & $k_-$ 
         & $\frac {k_-}{k_-+k_+}\times \frac13$ 
         & $\frac{k_-}{3K}$\\
\end{tabular}
\end{table*}

Then processes like $A00\to AAA\to 00A$ result in an {\em effective
  diffusion} for particle $A$ with a hopping rate $r_{\rm hop}=k_-/2$
($k_-$ is the rate of the first transition and $1/2$ the branching
probability for the second) and an effective step width of two lattice
sites.  The hopping rate is different for the last step before two
particles annihilate ($A00A\to 00AA$), namely $k_-/3$.  {\em Pair
  annihilation}, $AA \rightarrow 00$, occurs with a rate $k_+c$, or
$2k_-/3$ from the state $A00A$. {\em Pair creation}, on the other
hand, occurs mainly through the process $0000\to 00AA \to AAAA \to
A00A$; the corresponding rate is $k_-/3K$ per lattice site and hence
largely suppressed with respect to the annihilation process as long as
the particle concentration is far above its steady-state value.  These
processes are summarized in Tab.~\ref{tab_rates}.  Other processes
involve more particles and are of higher order in terms of a power
series in $K^{-1}$. They are negligible if the number of vacancies is
small, $n_A \ll 1$, and the binding constant is large, $K\gg 1$. In
summary, for $K \gg 1$, our dimer model can be mapped onto a
one-particle reaction-diffusion model $A+A \rightarrow 0$. Pair
creation processes, $0 \rightarrow A+A$, are highly suppressed during
the first stage of the relaxation process.  The give a significant
contribution to the dynamics close to the steady state where the
relaxation becomes exponential instead of algebraic.

Models like the simple reaction-diffusion model $A+A \rightarrow 0$
show interesting non-equilibrium
dynamics~\cite{privman97,mattis98,schuetz98}.  One can
show~\cite{torney83,toussaint83} that asymptotically the particle
density $n_A (t)$ decays algebraically, $n_A (t) \propto t^{-1/2}$,
which nicely explains the slow decay observed in simulation data (see
Fig.\ \ref{fig_relax}).  Note that a mean-field like rate equation
\begin{equation}
\partial_t n_A(t) \propto - n_A^2 (t) 
\end{equation}
would predict $n_A (t) \propto t^{-1}$. In our analysis we can even go
beyond the asymptotic scaling analysis and try to compare with exact
solutions of the model for a random initial distribution with density
$p$ by Krebs {\it et al.}~\cite{krebs95}.  They find (adapted to our
situation with two-site hopping)
\begin{eqnarray}
\label{eq_rd}
  n_A (t) = \frac{1}{2\pi}
  \int\limits_0^2 du \, 
  \frac{ \sqrt{u(2-u)} \, \left(n_0^{\rm plateau}\right)^2 \, 
         e^{-16u r_{\rm hop} t}}
       {u \, \left( u (\frac12-{n_0^{\rm plateau}}) + 
             \left( n_0^{\rm plateau}\right)^2 \right)} \;.
\end{eqnarray}
Its asymptotic limit (first determined by Torney and
McConnel~\cite{torney83}) reads 
\begin{equation}
\label{eq_torney}
n_A=(32\pi r_{\rm hop} t)^{-1/2}
\end{equation}
Note that it is independent of the initial particle concentration in
the Flory plateau ${n_0^{\rm plateau}}$. 

Our Monte-Carlo data (see Fig.\ \ref{fig_relax}) are in excellent
agreement with the predictions of Eq.\ \ref{eq_rd}. Minor deviations
at times between the plateau and the power-law decay are due to the
assumption of a random particle distribution underlying the derivation
of Eq.\ \ref{eq_rd}.  In reality the state after initial binding (RSA)
contains some correlations which however do not reach beyond a vacancy
(Markov-shielding) \cite{evans93}.  In other words, the sizes of
occupied areas are not exponentially distributed, but they are
uncorrelated to each other and the higher-order correlation functions
decay as fast as 2-point correlations (super-exponentially).  Short
ranged correlations in the initial configuration do have some effect
on concentration at intermediate times and this causes the deviation
between the simulation data and the theoretical curve which, however,
is not large.  According to \cite{family91,santos97} correlations can
affect the concentration in the asymptotic limit even if the pair
correlation function is short ranged and the order-$n$ correlation
functions decay at a length-scale of $\mathcal{O}(n)$, but this is not
the case in our model.

The results from the $A+A \rightarrow 0$ model also become invalid for
very long times where the particle concentration comes close to its
equilibrium value. In this limit the dynamics becomes scale-invariant
\cite{racz85}. The particle concentration can be written in scaling
form
\begin{equation}
\label{eq_scaling}
n_A(t)=K^{-1/2} \hat n( k_- t /K)
\end{equation}
with a diverging characteristic time scale $\tau_K \propto K$.

\subsubsection{Steady-state autocorrelation functions}
\label{subsec_steadystate}

\begin{figure}
\centerline{\includegraphics{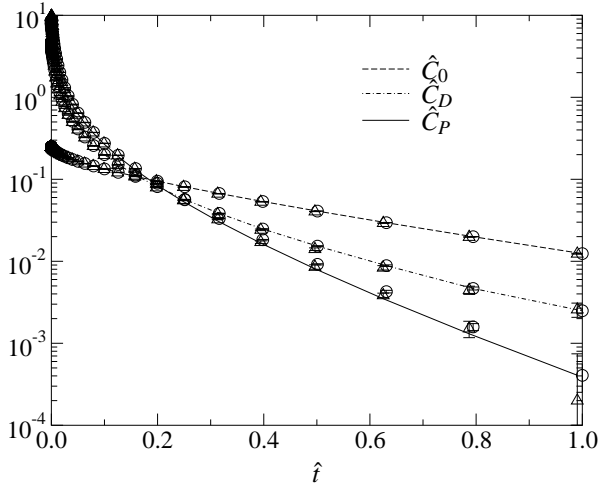}}
\caption{\label{d_fig_corr}Scaled equilibrium autocorrelation functions for 
  $\hat C_0(\hat t)=K C_0(t)$, $\hat C_D(\hat t)= C_D(t)$ and $\hat
  C_P(\hat t)=K C_P(t)$ , $\hat t=k_- t /K$. The function $\hat
  C_P(\hat t)$ is compared with the analytical result
  (\ref{d_eq_cpscale}), represented by the solid line.  Simulation
  data were obtained with $K=100$ (circles, for $C_0$ and $C_D$
  connected with dot-dashed and dashed lines) and $K=400$
  (triangles). \newline \tiny \copyright~2001, EDP Sciences \cite{vilfan2001b}}
\end{figure}

A limitation of the above mapping becomes evident if one considers the
equilibrium autocorrelation functions.  Contrary to conventional
$A+A\leftrightarrow 0$ models there are three different
autocorrelation functions with different functional forms (Fig.\ 
\ref{d_fig_corr}).  In the following we will use $\hat n _{( i, i+1
  )}$ as the occupation number (0 or 1) of a dimer on the pair of
sites $(i,i+1)$ and $\hat n_{0_{i}}$ as the occupation number of a
vacancy at the site $i$ (1 if there is a vacancy, 0 otherwise).  The
\textit{dimer autocorrelation function}
\begin{equation}
C_D(\tau)=\left< \hat n _{( i, i+1 )} (t)
  \hat n _{( i, i+1 )} (t+\tau) \right> - \left<  \hat n _{( i, i+1 )} (t)
  \right> ^2
\end{equation}
describes the probability to find a dimer at a given pair of sites
 simultaneously at times $t$ and $t+\tau$.  Similarly we
define the \textit{vacancy autocorrelation function}
\begin{equation}
C_0(\tau)=\left< \hat n_{0_{i}} (t) \hat n_{0 _{i}} (t+\tau) \right>
- \left< \hat n_{0_{i}} (t) \right>^2
\end{equation}
describing the probability that the site $i$ is vacant at times $t$ and $t+\tau$.
To calculate their values at $\tau=0$, we need the expectation values
\begin{equation}
\left< \hat n _{( i, i+1 )} (t) \right> =
\left< \hat n^2 _{( i, i+1 )} (t) \right> = n_D=
\frac 1 2 \left( 1 - \frac 1 {\sqrt{4K +1}} \right)
\end{equation}
and 
\begin{equation}
\left< \hat n_{0_{i}} (t) \right> =
\left< \hat n^2 _{0_{i}} (t) \right> = n_0=
\frac 1 {\sqrt{4K +1}}\;.
\end{equation}
A third way of defining an autocorrelation function is to ask for the
probability that a vacancy is either on the site $i$ or on the site
$i+1$
\begin{multline}
C_P(\tau)=\Bigl< (\hat n_{0_{i}}+\hat n_{0_{i+1}}-\hat n_{0_{i}}\hat
  n_{0_{i+1}}) (t) \times \\ \times 
  (\hat n_{0_{i}}+\hat n_{0_{i+1}}-\hat n_{0_{i}}\hat n_{0_{i+1}}) (t+\tau)
  \Bigr>\\
  - \left< (\hat n_{0_{i}}+\hat n_{0_{i+1}}-\hat n_{0_{i}}\hat n_{0_{i+1}})
    (t) \right>^2\;,
\end{multline}
with
\begin{equation}
\left< \hat n_{0_{i}}+\hat n_{0_{i+1}}-\hat n_{0_{i}}\hat n_{0_{i+1}} \right>=
\frac {2n_0}{1+n_0}=\frac 2 {\sqrt{4 K +1} +1}\;.
\end{equation}
The important difference between $C_P$ and $C_0$ originates from the
fact that the vacancies diffuse by hopping over two lattice sites.
This is why a vacancy diffusing along odd sites will not influence the
function $C_0$ if the original vacancy was on an even site.  The
function $C_P$ does not make this distinction.

For $K\gg 1$ the autocorrelation functions become scale invariant as
well.  Their scaling form reads $\hat C_0(\hat t)=K C_0(t)$, $\hat
C_D(\hat t)= C_D(t)$ and $\hat C_P(\hat t)=K C_P(t)$ with $\hat t=k_-
t /K$. The latter corresponds to the autocorrelation function in a
reaction-diffusion model $A+A\leftrightarrow 0$, which has recently
been calculated analytically by Bares and Mobilia \cite{bares99}
\begin{multline}
\label{d_eq_cpscale}
\hat C_P(\hat t)=\left( \frac {e^{-2 \hat t}}{\sqrt {2 \pi \hat t}} -
  \mbox{Erfc}\,\sqrt{2 \hat t} \right)  \mbox{Erfc}\,\sqrt{2 \hat t}\;, \\
  {\rm with}\quad \mbox{Erfc}\,(x)=\frac 2 {\sqrt{\pi}} \int_x^\infty
  e^{-y^2} dy\;.
\end{multline}
Note that the described solution is derived for a parameter set which
requires a special relation between diffusion, pair-creation and
annihilation rate to be fulfilled. In our model this is not the case.
But, this difference becomes irrelevant in the scaling limit, since
their and our model can be mapped onto each other by introducing a
short-ranged interaction between particles. See also
Ref.~\cite{park01} for a comment about the validity of the analytical
approximation used in Ref.~\cite{bares99}. The observed deviations
however do not affect the scaling limit.

The other two functions decay on the same time-scale, but with
different prefactors. The reason is that even if a pair of vacancies
annihilates, the system still keeps memory on whether the surrounding
dimers were located on even or odd locations and this gives those
correlation functions that distinguish between even and odd sites a
longer decay time.

\subsection{Two-particle model}

Another interesting case arises if we consider the limit $K_2 \to
\infty$ with $K_1 c$ being of the order of magnitude $1$.  In this
case neither the dimers bound with one head nor the empty lattice
sites dominate and we have to introduce two particle types.  As we are
used to, particles $A$ should denote vacancies on the lattice.  In
addition, we introduce particles $B$, representing dimers bound with
only one head.

\begin{figure}
\begin{center}
\includegraphics{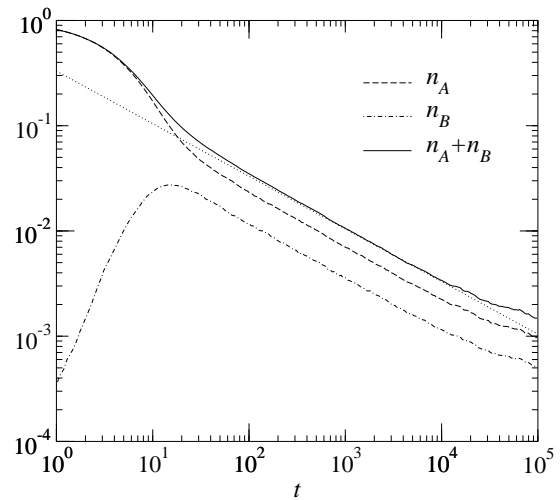}
\end{center}
\caption{\label{d_fig_ab} A computer simulation showing the time dependence of
  the vacancy ($n_A$) density, the single-bonded dimer ($n_B$) density
  and their sum for $K_1c=0.5$ and $K_2=10^6$.  The dotted line
  shows the predicted long-time limit of reaction-diffusion models
  given by Eq.~(\ref{eq_torney}).  In detail, parameters are:
  $k_{+1}c=0.1$, $k_{-1}=0.2$, $k_{+2}^f=10^6$, $k_{-2}^f=1$,
  $k_{+2}^b=0.1\, k_{+2}^f$ and $k_{-2}^b=0.1\, k_{-2}^f$.  The
  simulation has been performed on a lattice with $2^{14}$ sites and
  periodic boundary conditions.}
\end{figure}

Reactions $A\to B$ and $B\to A$ occur at rates $k_{1+}c$ and $k_{1-}$.
If we assume that both attachment rates of the second head $k_{2+}^f$
and $k_{2+}^b$ are significantly higher than $k_{1+}c$ and $k_{1-}$,
we are again dealing with the \emph{diffusive model}
(\ref{eq_diffusive_condition}).  The diffusion mechanism for vacancies
(particles A) is exactly the same as in the model with stiff dimers
and according to (\ref{d_dif}) the hopping rate is given by
Eq.~(\ref{d_dif}).  The diffusion rate of single-bonded dimers
(particles B) has to be equal.  This is due to the symmetry of our
model upon exchanging particles A and B and the transition rates
$k_{1+}c$ and $k_{1-}$ (Tab.~\ref{tab_symmetries}) and because the
latter are irrelevant for the hopping rate.  If a particle A and a
particle B reach neighboring sites, they annihilate quickly, while
two particles of the same type do not interact.  To summarize, we
obtain a reaction diffusion model of the type:
\begin{align*}
A&\leftrightarrow B \\
A+B&\leftrightarrow 0
\end{align*}
While its dynamics is more complex at short times, the model becomes
equivalent to the $A+A\leftrightarrow 0$ if the transitions between
$A$ and $B$ become faster than the typical annihilation time, which is
given as the diffusion time between two particles, displaced by the
average distance between particles on the lattice, $\sim
\frac{1}{r_{\rm hop} n^2}$.  Therefore the power-law behavior, given
by (\ref{eq_torney}) remains untouched by the fact that we are dealing
with two different particle types.  An example of a simulation in this
regime is shown in Fig.\ \ref{d_fig_ab}.

\subsection{Dynamics of interacting dimers in one dimension}
\label{sec:dynam-inter-dimers}

\begin{figure}[thbp]
\centerline{\includegraphics{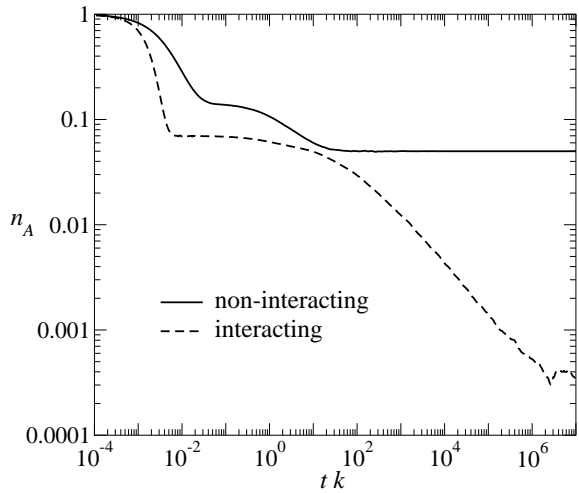}}
\caption{The number of vacant lattice sites as a function of time 
  non-interacting (solid line) and interacting (dashed lines) dimers.
  The parameters were $K=100$, $A=10$ and $B=1/10$ (a dimer is 10
  times as likely to associate to and 10 times less likely to
  dissociate from a certain site if one of the neighbors is present)}
\label{fig_interacting_relax}
\end{figure}

In many biological systems the interaction between attached molecules
plays an important role.  For example, in the case of kinesin, there
has been an observation of coexisting empty and decorated domains
which can only be explained by an attractive interaction
\cite{vilfan2001a}.  Similar observation has been made on actin
decorated with double-headed myosin \cite{orlova97}.  An interaction
between the dimers can be introduced by assuming that a dimer is more
likely to bind to a pair sites if one or two neighbors are already
bound.  The binding rate then becomes $k_+^P=k_+^N=A k_+^N$ (one
neighbor bound) or $k_+^D=A^2 k_+^N$ (both neighbors bound).
Similar, we assume that a dimer with one bound neighbor dissociates
with rate $k_-^P=k_-^M=B k_-^N$ and that a dimer with two bound
neighbors dissociates with rate $k_-^D=B^2 k_-^N$.  The equilibrium state of
this model is still exactly solvable \cite{mcghee74}, but the interaction
changes both relaxation stages quantitatively.  First, the vacancy
concentration on the intermediate plateau lowers since the interaction
improves the formation of contiguous clusters during the first stage.
The RSA phase can then be described in terms of Kolmogorov's
grain-growth model \cite{kolmogorov37}.  The vacancy concentration
after the initial binding is given as $n_0 \approx \frac12
\sqrt{\frac{\pi r_{\rm nuc}}{2 r_{\rm growth}}}$
\cite{gonzalez74,evans83}, where $r_{\rm nuc}$ is the nucleation rate
per free lattice site (in the simplest case, when $K \gg 1$, it is
simply $k_+$) and $r_{\rm growth}$ the sum of speed with which both
boundaries of a nucleus spread over the lattice (normally
$k_+^P+k_+^M=2 A k_+^N$).  The second effect of the interaction is
that the diffusional relaxation slows down by the factor $B$ since the
detachment rate decreases.  And finally, the equilibrium vacancy
concentration decreases \cite{mcghee74}.  Nevertheless, interacting
models show the same two-stage relaxation behavior.  An example of a
model with interaction is shown by the dot-dashed line in Fig.\ 
\ref{fig_interacting_relax}.

\section{Dimer model with interaction}
\label{sec:dimer-model-with}

As we mentioned in Section\ \ref{sec:dynam-inter-dimers}, interactions
between bound dimers can often play an important role.  In the
kinesin-microtubule system, evidence for the existence of an
attractive interaction comes from two key experimental observations.
The first one is that bound kinesin dimers are found to form
two-dimensional crystalline lattices~\cite{thormaehlen98}.  Within the
non-interacting model one can explain longitudinal correlations along
a single protofilament, but by no means the alignment of dimers on
different protofilaments.  The second observation supporting an
attractive interaction between bound dimers is the coexistence of
empty and decorated microtubules in one and the same
experiment~\cite{vilfan2001a}.  Such a phase segregation is a clear
indication that the system is in a coexistence regime with the
strength of the dimer interaction above some critical value. The
detailed nature of the interaction between the dimers is not yet
known. It could be a direct interaction between kinesin heads and
tails or some indirect interaction mediated through distortions of the
underlying tubulin lattice or a combination of all of these
possibilities.  All of these mechanisms are consistent with the
observation that the interaction is stronger on flat tubulin sheets
than on cylindrical microtubules \cite{vilfan2001a}.

\begin{figure}[thbp]
\begin{center}
  \includegraphics{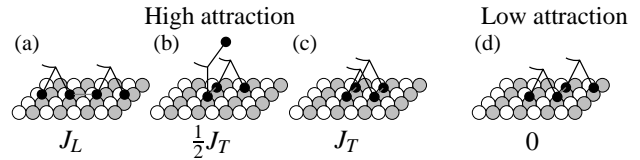}
\end{center}
  \caption{Illustration of the interaction energy between two 
    adsorbed dimers in different relative positions. \newline \tiny \copyright~2001, Academic Press \cite{vilfan2001a}}
  \label{fig_interactions}
\end{figure}

Even if one assumes that there is only nearest-neighbor interaction, a
general description would still require $12$ different coupling
strengths ($2$, $4$ and $6$ for describing the attraction between two
single-bonded dimers, two double-bonded dimers and between a single-
and a double-bonded dimer, respectively).  Upon neglecting that the
tubulin lattice is actually rhombic and not orthogonal one can reduce
the number of different coupling constants to $9$. These are still too
many for a general discussion.  Hence, mainly for simplicity, we will
restrict ourselves to a model with only two interaction constants,
$J_L$ acting in longitudinal and $J_T$ acting in transverse direction,
as illustrated in Fig.\ \ref{fig_interactions}.  We assume that the
interaction between two double-bonded dimers on neighboring
protofilaments is zero if they are displaced by one lattice site or
more (Fig.\ \ref{fig_interactions}d).  Otherwise, the transverse
interaction is assumed to have the strength $J_T$ between two
double-bonded dimers (Fig.\ \ref{fig_interactions}c) and $\frac12 J_T$
between a double- and a single-bonded dimer (Fig.\ 
\ref{fig_interactions}b).  We use the same notation as in Sec.\ 
\ref{sec:dynam-inter-dimers} and introduce factors $A_i$ and $B_i$
measuring the change in the attachment and detachment rate by the
presence of a neighboring dimer (with $i$ indicating the relative
position of the neighbors as shown in Fig.\ \ref{fig_interactions}),
\begin{align*}
  k_{1,2+}^{\text{with neighbour}} &= 
  A_i\, k_{1,2+}^{\text{without neighbour}} \; ,
\\
  k_{1,2-}^{\text{with neighbour}} &= 
  B_i\, k_{1,2-}^{\text{without neighbour}} \; .
\end{align*}
Detailed balance states that
\begin{equation}
J_i=k_B T \ln \frac {A_i} {B_i}\;.
\end{equation}
For simplicity we further assume that the binding and the unbinding
rate are always changed by the same factor, $A_i=1/B_i$; this
assumption is irrelevant in equilibrium, but it simplifies the
kinetics.

\begin{figure}[thbp]
\begin{center}
  \includegraphics*{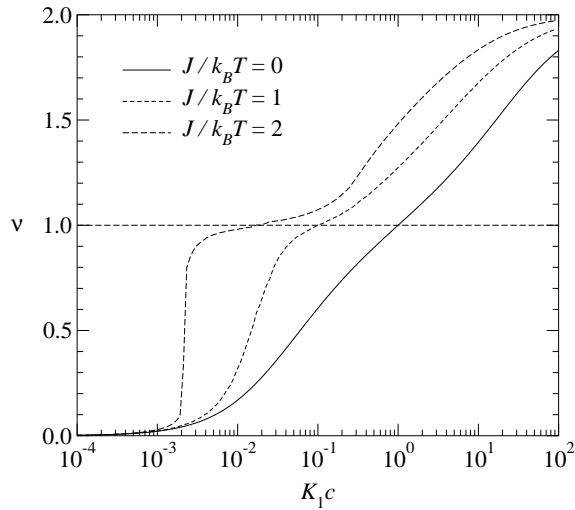}
\end{center}  
  \caption{Binding stoichiometry of the interacting dimer model (dashed
    lines) compared with the non-interacting (solid line). The binding
    constant of the second head is $K_2=10$, the interaction strength
    $J=J_T=J_L=k_BT$ (short-dashed line) and $2\,k_B T$ (long-dashed
    line).}
  \label{fig_stoch_interacting}
\end{figure}

Fig.\ \ref{fig_stoch_interacting} shows the stoichiometry curves,
equivalent to those in Fig.\ \ref{fig_curves}, for the interacting
model.  The most dramatic effect of the interaction is that it
broadens the plateau where most dimers are double-bonded, giving a
stoichiometry of one head per lattice site.  If the interaction is
strong enough (see e.g. the curve with $J/k_BT=2$ in Fig.\ 
\ref{fig_stoch_interacting}), the stoichiometry curves show quite a steep rise
from zero to the plateau at $\nu = 1$. This is quite indicative of a
phase transition, albeit smeared somewhat out due to finite-size
effects.

To determine the critical point we make two further simplifications.
First, we assume that we are in the stiff dimer limit, $K_2 \gg 1$, as
we did in Sec.\ \ref{sec:stiff-dimer-limit}.  This stiff dimer model
can be interpreted as a spin model with \emph{axial next nearest
  neighbor interaction}, also known as ANNNI models, reviewed by Selke
\cite{selke88}. In terms of these ANNNI models, our model is a rather
peculiar limit since it has an infinitely repulsive interaction
between nearest neighbors and attractive interaction between
next-nearest neighbors in longitudinal direction, as well as an
attractive interaction between nearest neighbors in transverse
direction.  In the literature on ANNNI models mainly the opposite
situation was studied since it leads to frustrations, while the
``zero-temperature'' state of our model is trivial, namely dimers
aligned, let us say, the even sites on the protofilaments.  This
justifies an approximation which further simplifies the model, namely
by assuming that dimers can only bind to even sites. Then the problem
simplifies to a lattice-gas model with an effective Hamiltonian
\begin{equation}
\mathcal{H}=\sum_{i,j}\left( -J_L \hat n_{2i,j} \hat n_{2i+2,j} 
-J_T  \hat n_{2i,j}  \hat n_{2i,j+1} -\mu \hat n_{2i,j} \right) \; ,
\end{equation}
where the coordinates $(2i,j)$ run over all even lattice sites, $\hat
n_{2i,j}$ are the occupancy numbers ($0$ or $1$) on those sites and
$\mu=k_B T \ln (K_1 K_2 c)$ is the chemical potential of adsorbed
dimers.  The lattice gas model can be mapped onto the 2D Ising model
\cite{yeomans92} with Hamiltonian 
\begin{equation}
\mathcal{H'}=\sum_{i,j}\left( -J'_L s_{i,j} s_{i+1,j} 
-J'_T  s_{i,j}  s_{i,j+1} -\epsilon s_{i,j} \right) 
\end{equation}
with exchange constants 
\begin{equation}
\label{eq_transformtoising_exchange}
J_T'=\frac14 J_T \, ,\qquad  J_L'= \frac14 J_L \, ,
\end{equation}
and spin variables
\begin{equation} 
\label{eq_transformtoising_spin}
s_{i,j}=2\hat n_{2i,j}-1 \; .
\end{equation}
The spin $s$ has the value $+1$ if it is parallel and $-1$ if it is
antiparallel to the magnetic field $\epsilon$,
\begin{equation}
\label{eq_transformtoising_field}
\epsilon=\frac {\mu +J_L+J_T}2\;.
\end{equation}
  
The critical temperature of the two dimensional Ising model was first
determined by Kramers and Wannier \cite{kramers41}.  In the absence of
an external field, the model has also been exactly solved by Onsager
\cite{onsager44}.  The condition for the critical point reads
\begin{equation}
\sinh \frac {J_L}{2k_BT}\times \sinh \frac {J_T}{2k_BT}=1\;, 
\end{equation}
or $J_c=1.76\, k_BT$ in the isotropic case ($J_L=J_T$).

The exact solution helps us to determine the critical point, {\it
  i.e.}\/ the interaction strength needed to obtain a phase transition
between an empty and a decorated state.  The full form of the binding
stoichiometry as a function of concentration, however, would be
equivalent to calculating the magnetization of the two-dimensional
Ising model in the presence of an external field, which has not yet
been achieved analytically.  Therefore, one has to rely on computer
simulations to obtain the stoichiometry curves.  An overview on
Monte-Carlo simulations on lattice gas models can be found in
Ref.~\cite{binder89}.  The situation simplifies when one is far away
from the critical point, {\it i.e.}\/ when the coupling is much
stronger than its critical value, $J\gg J_c$.  This corresponds to the
low-temperature limit in the Ising analogue, where the spin
polarization is saturated and points into the direction of the
external field. The number of spins with orientation $\uparrow$ is
given as $n=N \Theta(\epsilon)$, where $\Theta$ is the Heaviside
function.  Back in the adsorption model, according to Eqs.\ 
(\ref{eq_transformtoising_exchange})-(\ref{eq_transformtoising_field}),
the condition reads
\begin{equation}
\label{eq_highj}
\nu = \Theta(\mu+J_L+J_T)\;.
\end{equation}
This expression can also be understood directly. The boundary of a
decorated domain will normally consist of many edge elements (each one
with three out of four neighboring sites occupied) and some corner
elements (with two out of four neighboring sites occupied).  If the
corner elements are stable, the decorated domain will grow; if they
are unstable, it will shrink.  The stability condition of a corner
element with one bound longitudinal neighbor and one bound transverse
neighbor is given by Eq.~(\ref{eq_highj}).  Deviations from
(\ref{eq_highj}) are possible due to finite size effects. This is
because small patches have a higher curvature in the boundary and are
therefore less stable.  

\subsection{Dynamics of the interacting 2D model}

We have shown in Sec.\ \ref{sec:power-law-regime} that the annealing
of gaps between bound dimers can be slowed down substantially as
compared to the transition rates for single dimers.  This effect is
even stronger in a two dimensional system.  Here linear domain
boundaries emerge instead of point defects (gaps).  These boundaries
can annihilate when two of them join.  But their diffusion is
significantly slower than the diffusion of single gaps, roughly
inverse proportional to the number of sites in the boundary.

The nonequilibrium dynamics of decorating an initially empty
two-dimensional lattice is an interesting and complex problem. It has
first been studied by Becker and D{\"o}ring~\cite{becker35} (reviewed
in Ref.~\cite{lewis78}). A crucial concept is the critical nucleus, a
certain number of adsorbed atoms (here dimers) necessary to form a
stable nucleus able to grow and spread over the whole lattice.  An
analytical expression for the nucleation time can be given if the
interaction is strong enough, and the critical cluster size becomes as
small as four, three or two molecules. For larger critical cluster
sizes a rough estimate says that the nucleation rate is proportional
to the kinesin concentration to a power that is typically half the
number of nucleus-forming units.

\begin{figure*}[!t]
\begin{center}
\begin{tabular}{l@{\hspace{2cm}}l}
\begin{tabular}[b]{l}
\includegraphics{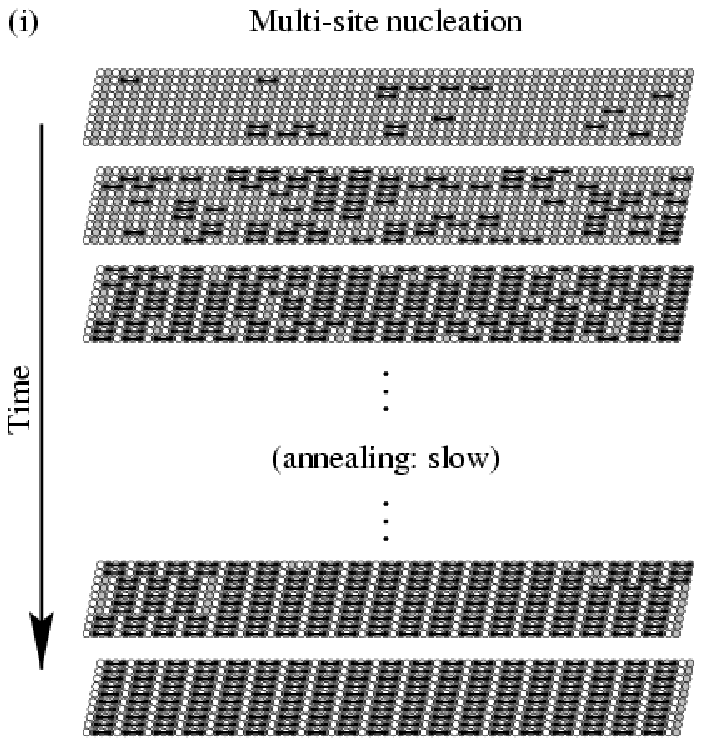}\\[1cm]
\includegraphics{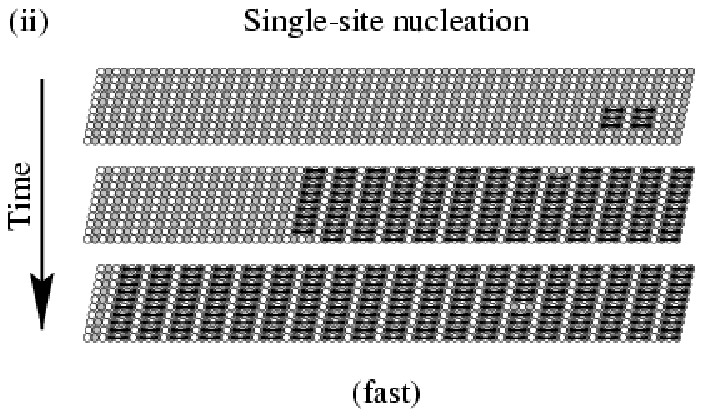}
\end{tabular}&
\begin{tabular}[b]{l}
\includegraphics{Figure15b.eps}
\end{tabular}
\\
a)&b)
\end{tabular}
\caption{\label{fig_sepnucphase} a) Computer simulation
  of (i) multiple-site nucleation with subsequent annealing and (ii)
  single-site nucleation. Both processes can lead to a defect-free
  decoration.  However, annealing can be extremely slow whereas
  single-site nucleation immediately leads to the ordered state. The
  parameters in the simulation were the following: (i): $K_1 K_2
  c=0.1$, $A_L=A_T=3$, $B_L=B_T=0.3$ (ii): $K_1 K_2 c=0.001$,
  $A_L=A_T=10$, $B_L=B_T=0.1$. b) The solid line shows the phase
  transition above which a decorated phase exists. The dashed line
  shows the border between homogeneous (defect-free) decoration
  reached immediately through nucleation and decoration with domain
  walls. The curves were obtained from a computer simulation on a
  lattice of $14\times100$ sites (note that the defect-free range
  would be somewhat larger on a smaller lattice). \newline \tiny \copyright~2001, Academic Press \cite{vilfan2001a}}
\end{center}
\end{figure*}

There are different scenarios depending on the relative magnitude of
the time to form a critical nucleus, $t_{\rm nuc}$, and its growth
time, $t_{\rm growth}$, {\it i.e.}\/ the time it takes such a nucleus
to grow and cover the whole surface (see Fig.\ 
\ref{fig_sepnucphase}a).\footnote{For a Java applet showing the
  nucleation in the two-dimensional dimer model see {\tt
    http://www.ph.tum.de/\~{}avilfan/decor/}~.}  If the nucleation time
is larger than the growth time,
\begin{equation}
\label{eq_nuc_vs_growth_a}
t_{\rm nuc} \gtrsim t_{\rm growth}\;,
\end{equation}
the whole lattice will most probably be completely covered with dimers
before other nucleation sites can form anywhere else on the lattice.
The final state is a defect free lattice generated from a single
nucleus.  This is consistent with experimental observations
\cite{thormaehlen98}. At the same time single-site nucleation also
explains how empty and fully decorated microtubules can be found
coexisting in one and the same experiment~\cite{vilfan2001a}.  When
the decoration starts, nuclei will form on certain microtubules and
totally decorate them, until the concentration of free kinesin in
solution drops below the critical value, where the equilibrium phase
is the empty lattice.  A numerical study of the nucleation
process~\cite{vilfan2001a} gives us an estimate of the minimum
interaction strength needed to fulfill the condition
(\ref{eq_nuc_vs_growth_a}) as $J\approx 3\, k_B T$. This strongly
indicates that there is a interaction between kinesin dimers which is
of the order of $3\, k_B T$.
If the nucleation time is smaller than the growth time,
\begin{equation}
\label{eq_nuc_vs_growth_b}
t_{\rm growth} \gtrsim t_{\rm nuc}\;,
\end{equation}
there is a certain probability that a second critical nucleus forms on
the lattice before the first one created had a chance to cover the
whole lattice. The probability of such or higher order events grows
with decreasing nucleation time. Then, the lattice will be covered by
several domains, locally ordered areas which may be out of phase on
the microtubule lattice. In order to finally reach a homogeneous
decoration with only one predominant domain a secondary process is
needed which leads to a coarsening of the initial grain-like structure.
This secondary process includes domain wall wandering and annihilation
induced by the detachment and re-attachment of kinesin dimers.  As
mentioned above, such annealing processes are expected to be much
slower than the unhindered growth of individual domains. Fig.\ 
\ref{fig_sepnucphase}b summarizes the possible phases and dynamic
regimes for interacting dimers decorating a two-dimensional lattice.

\section{Interacting $k$-mers in one dimension}
\label{sec:interacting-k-mers}

Non-interacting $k$-mers ($k>3$) are well described with the
mean-field kinetics of the model $kA\to 0$ which predicts a gap
density decay $\propto t^{-1/(k-1)}$ \cite{nielaba92,privman92a}.  In
the continuous limit, also called continuous sequential adsorption
(CSA) or ``car parking problem'' the length of unoccupied space decays
as $\propto 1 / \log t$ in the mean-field
regime~\cite{krapivsky94,jin94}.

The dynamics becomes more intriguing when a strong attraction between
the $k$-mers is introduced.  An example of such a system is
tropomyosin~\cite{vilfan2001c}, a protein that plays a key role in the
regulation of muscle contraction.  In muscle each tropomyosin molecule
binding on an actin filament occupies seven actin monomers (although
other isoforms covering 5 or 6 monomers exist as
well~\cite{hitchcock-degregori94}) and thereby obstructs the action of
myosin motors.  The binding of tropomyosin to actin is strongly
cooperative -- the binding constant next to a bound molecule is 1000
times larger than at an isolated place~\cite{wegner79}.

\begin{figure}[thbp]
\begin{center}
\includegraphics{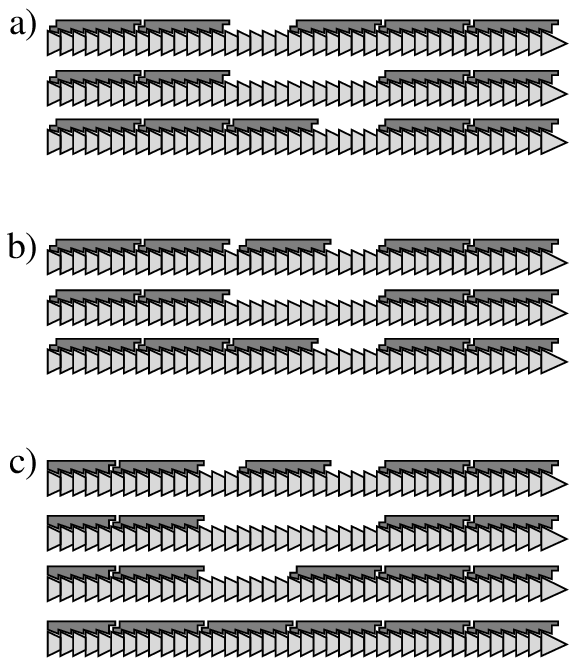}
\end{center}
\caption{Examples of effective reactions that
occur after the detachment and attachment of a heptamer:
Diffusive step (a); Pair coagulation $A+A\to A$ (b); Pair annihilation
$A+A\to 0$ (c). \newline \tiny \copyright~2001, Biophysical Society \cite{vilfan2001c}}
\label{fig_effective_steps} 
\end{figure}

The initial binding phase is analog to the problem of interacting
dimers and can again be described using a grain growth
model~\cite{kolmogorov37,evans93,vilfan2001c}.  After the initial
phase, gaps with sizes between $1$ and $k-1$ sites will remain.  Their
sizes are distributed randomly with equal probabilities of $1/(k-1)$
for each gap size. What follows is again a process on a much longer
time-scale in which $k$-mers at edges of the gaps can detach and
re-attach.  If they detach from one side of the gap and then
re-attach at the other side, the gap moves by $k$ sites in one
direction (see Fig.\ \ref{fig_effective_steps}a).
The rate of such diffusive steps is given as the detachment rate of a
molecule on e.g.\ the minus side of a block multiplied by the
probability that a molecule re-attaches on the plus side of the
neighboring block before another $k$-mer can re-attach to the position
where the first one has detached:
\begin{equation}
\label{eq_hop}
r_{\rm hop}=k_-^M \frac {k_+^P}{k_+^P+k_+^M}=
k_-^M \frac {k_-^P}{k_-^P+k_-^M}= \left( \frac 1 {k_-^P}+\frac 1 {k_-^M}
\right)^{-1} \;.
\end{equation}
Two neighboring gaps (see Figs.\ \ref{fig_effective_steps}b and c),
each containing between $1$ and $k-1$ sites with randomly distributed
probabilities, can either join to a single gap (Fig.\ 
\ref{fig_effective_steps}b) or, if their total size exactly fits one
$k$-mer, annihilate (Fig.\ \ref{fig_effective_steps}c).  If the
original gap sizes are $g_1$ and $g_2$, the joined gap has size
$(g_1+g_2) \mod k$. The annihilation will take place in $k-1$ out of
$(k-1)^2$ cases, therefore its probability is $1/(k-1)$.  Any other
gap size will be the outcome in $k-2$ cases, therefore having the
probability $(k-2)/(k-1)^2$.

Again we can represent gaps as particles $A$ hopping randomly along
the lattice and joining or annihilating when two of them meet.  The
pair creation processes become relevant only if the particle
concentration approaches its equilibrium value and will be neglected
for now.  Then, we can map our model to a
diffusion-coagulation-annihilation model consisting of following
reactions
\begin{align*}
  A+A&\to A& \text{with probability}&\; (k-2)/(k-1) \;,\\
  A+A&\to 0& \text{with probability}&\; 1/(k-1) \;.
\end{align*}
The fundamental difference between the interacting $k$-mer and the
dimer model is that the $k$-mer model not only contains pair
annihilation but also pair coagulation processes, $A+A\to A$.
According to Ref.~\cite{lee94} all diffusion-annihilation models of the
type $2A\to l A$ (or generally $m A\to l A$) belong to the same
universality class. In the asymptotic limit, the ``particle''
concentration can then be borrowed from the exactly solvable $A+A\to
0$ model~\cite{lushnikov87,torney83}. Adapted to our $k$-mer model it
reads
\begin{equation}
n_A(t)=\frac 2 {2-l} \frac 1 {\sqrt{8 \pi \bar{r}_{\text{hop}} t}}
\end{equation}
with $l=(k-2)/(k-1)$ and $\bar{r}_{\text{hop}}=k^2 r_{\text{hop}}$;
the second relation results from the fact that in each diffusive step
a gap jumps over $k$ sites. Hence, we finally obtain
\begin{equation}
\label{eq_ngt}
n_A(t)=\frac {k-1} {k^2 \sqrt{2 \pi r_{\text{hop}} t}} \; .
\end{equation}
Interestingly, the asymptotic particle concentration is independent of
its initial value, {\it i.e.}\/ the final gap concentration at long
times does not depend on the intermediate gap concentration $n_A^0$.
Note that it is even independent of the solution concentration $c$.

\begin{figure}[thbp]
  \centerline{\includegraphics{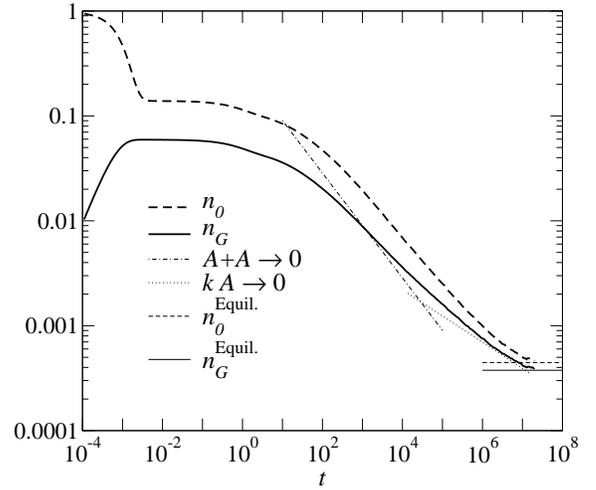}}
  \caption{The time-dependent gap concentration ($n_G$) and fraction
    of empty lattice sites ($n_0$) for an interacting pentamer
    ($k=5$), as obtained from the simulation.  The dot-dashed line
    shows the prediction of Eq.~(\ref{eq_ngt}) and the dotted line the
    mean-field power law $\propto t^{-1/(k-1)}$.  The thin solid and dashed
    line show the equilibrium values of $n_G$ and $n_0$ using the expression
    from Ref.~\cite{mcghee74}. The parameters are:
    $k_+^N c=100$, $k_+^P c=k_+^M c=1000$, $k_-^N=1$, $k_-^P=k_-^M=0.1$,
    $k_+^D c=10000$ and $k_-^D=0.01$.}%0.000445
  \label{fig_pentamer}
\end{figure}

Of course, the mapping to the $A+A\to l A$ model is valid only as long
as the concentration of particles $A$ is well above its equilibrium
concentration.  Otherwise, particle splitting events like $A\to A+A$
also become relevant.  If the concentration $c$ is high enough, the
$t^{-1/2}$ regime will be followed by a regime where, due to these
splitting events, the average gap size will finally approach one site,
which then are randomly distributed over the lattice.  The number of
gaps will decay according to the mean-field prediction $\propto
t^{-1/(k-1)}$, as it does in the non-interacting $k$-mer
model~\cite{nielaba92,privman92a}.  An example of this behavior is
shown in Fig.\ \ref{fig_pentamer}.  When the system approaches
equilibrium, processes of pair creation $0\to A+A$ become important as
well.  Similar to the dimer model, these processes leads to a final
exponential relaxation towards equilibrium.  The equilibrium gap size
and distribution of the interacting $k$-mer model are known exactly
from a study by McGhee and von Hippel~\cite{mcghee74}.

\section{Summary and Outlook}

In this contribution we have discussed the kinetics of some
macromolecular assembly processes relevant for the formation of
functional structures in cells. In particular, we were interested in
the dynamics of ligand-substrate binding, where the substrate is a
one- or two-dimensional lattice and the ligands are dimers or
oligomers.  As examples we picked binding of dimeric kinesin on
microtubules and tropomyosin on actin filaments. There are other
related systems where protein ligands bind to biological
macromolecules, e.g.\ proteins to DNA or antibodies to viruses.

Our theory allows us to study the effects of {\it steric constraints,
  binding rates and interaction between neighboring proteins} on the
binding dynamics and binding stoichiometry. Our key results are as
follows. The collective dynamics turns out to be not only much
slower than the kinetics of single molecules but it also shows some
interesting dynamic anomalies. Quite generally, we find that the
binding kinetics goes through several qualitatively different stages.
At small times when deposition processes are frequent but detachment
processes are still unlikely, the kinetics can be described in terms
of sequential deposition models. Such models have been analyzed in
various contexts years ago so that there are many exact results
available. After this initial phase the system is left in a
``metastable'' or ``locked'' configuration, \textit{i.e.} a
configuration with many small vacant regions between blocks of bound
molecules prohibiting binding of additional ligands to the substrate.
As a consequence one obtains an intermediate plateau regime for the
binding stoichiometry. For escaping from this locked configuration
vacant regions have to merge in order to accommodate for the deposition
of additional ligands. These are extremely slow processes mediated by
a sequence of detachment and re-attachment processes. Upon identifying
the gaps with particles $A$ we have shown that the dynamics in this
regime can be explained by mapping it onto reaction-diffusion models.
For stiff dimers the corresponding model is $A + A \rightarrow 0$
resulting in a gap density that decays with a power law $\propto
t^{-1/2}$ before it approaches the final equilibrium value. Taking
into account that actual biological dimers have some level of
flexibility the model had to be generalized to include a second
species of particles $B$, representing dimers bound with only one
head.  The corresponding reaction-diffusion model is then of the type
$A\leftrightarrow B$ and $A+B\leftrightarrow 0$.  While its dynamics
is more complex at short times, the model becomes equivalent to the
$A+A\leftrightarrow 0$ if the transitions between $A$ and $B$ become
faster than the typical annihilation time. If, on the other hand, one
introduces a strong diffusion of adsorbed dimers the power-law decay
is preceded by a mean-field regime where the gap density decays with
$\propto t^{-1}$. The two-stage relaxation and the mapping to the
$A+A\to 0$ model remain valid even if there is an attractive
interaction between adsorbed dimers on the one-dimensional lattice.
In the two-dimensional model with interaction, the sequential
deposition stage goes over into a nucleation process and the
relaxation stage into a much slower annealing process.  Yet even then
the two-stage relaxation remains qualitatively the same.

The annealing process for interacting $k$-mers with $k > 3$ is
fundamentally different from the dimer model. One finds that
non-interacting $k$-mers are always well described with the mean-field
kinetics of the model $k A \rightarrow 0$ predicting a decay in the
gap density $\propto t^{-1/(k-1)}$. Introducing a strong interaction
between the $k$-mers, as is the case for tropomyosin binding on actin
filaments, we find a mapping to the $A + A \rightarrow l A$ model,
valid for concentrations well above the equilibrium value, which shows
a particle concentration decaying $\propto t^{-1/2}$. Closer to
equilibrium other processes become important as well.  First, particle
splitting, $A\to A+A$ leads to an intermediate mean-field regime
similar to that of non-interacting $k$-mers. Later the events of pair
creation, $0\to A+A$ become relevant and lead the system into the
final exponential relaxation towards equilibrium.

All of the above results concern the statics and dynamics of ligands
decorating periodic structures in the absence of ATP hydrolysis. Such
systems are ``passive'' in the sense that the ligands bind and unbind
from their respective substrates but do not show any motor activity,
{\it i.e.}\/ move along these molecular tracks. As described above,
this allowed us to give a detailed description of decoration
experiments and to analyze corresponding experiments quantitatively.
We also were able to identify and quantify cooperative effects between
kinesin dimers.  As our results show, the relaxation time in many
experimental or biological systems can be extremely long.  Therefore,
one has to be aware that an experiment where the measurement is taken
soon after the start of the decoration often does not show the
equilibrium configuration, but rather some intermediate state from the
relaxation process.

A natural extensions of this work is to investigate ``active'' systems
composed of an ensemble of motor proteins and their respective
substrates in the presence of transport along the molecular tracks
driven by the chemical energy of ATP hydrolysis. Understanding such
systems lies at the heart of many cellular processes.  Although single
processive motors like kinesin or myosin V can move loads over
considerable distances, the number of motors acting simultaneously
when transporting vesicles through the cytoplasm can be pretty large.
For instance, in Ref.~\cite{feneberg2001} the force acting on a bead
in the cell has been estimated as $100-200\,{\rm pN}$, which implies
that at least $20-40$ motor proteins were involved.  How these active
motors cooperate at high densities and what role is played by
interactions between them, is to a large extend a still unexplored
field with many open questions.

An even broader range of questions arises when one tries to understand
how cells organize their interior to fulfill their various duties.
While organizing its interior, the cell has to physically separate
molecules or molecular aggregates from each other, define distinct
functional regions, and actively transport molecules between these
regions. Such processes rely on the assembly and ordering processes of
macromolecules inside the cell and on forces that may be generated by
molecular assembly or by the action of motor proteins.  Understanding
the principles underlying those complex cellular processes not only
requires biological and biochemical studies but also studies of
physical processes. In this respect microtubules and motors have been
used as {\em in vitro} model systems to study various aspects of
regulation and self-organization of cellular structures. It was shown
theoretically and experimentally that the dynamic instability of
microtubules in combination with microtubule polymerization forces is
sufficient to provide a microtubule organizing center with a mechanism
to position itself at the center of a confined geometry
\cite{holy97,dogterom97}.  In a simple {\em in vitro} system
consisting solely of multi-headed constructs of the motor protein
kinesin and stabilized microtubules the formation of asters and spiral
defects was shown \cite{nedelec97}. By varying the relative
concentration of the components a variety of self-organized structures
(patterns) where observed.  The pattern formation observed in these
simple {\em in vitro} systems are only first examples of what will be
a much broader range of self-tuned or regulated structural and
temporal organization in biological systems.  The control parameter
determining the structure was the relative concentration of the
components. In cellular processes there are, however, other possible
parameters for controlling and regulating such as binding of a host of
associated binding proteins. We expect that these and other related
biological processes are a source for fascinating cooperative effects
and a multitude of interesting nonequilibrium dynamic phenomena.

\begin{acknowledgments}
  We would like to thank Eckhard Mandelkow and Andreas Hoenger for
  helpful discussion about microtubule decoration experiments and
  their relevance for studying motor proteins. We have also benefited
  from discussions with Tom Duke, Thomas Franosch, Jaime Santos, Franz
  Schwabl, Gunter Sch{\"u}tz and Uwe C.  T{\"a}uber. Our work has been
  supported by the DFG under contract nos.\ SFB~413 and FR~850/4-1.
  A.V. would like to acknowledge support by the European Union through
  a Marie Curie Fellowship (contract no.~HPMFCT-2000-00522).
\end{acknowledgments}

% Bibliography 

\end{document}